\documentclass[aps, reprint,showpacs,showkeys,pra,superscriptaddress]{revtex4-2}
\usepackage{dcolumn}
\usepackage[utf8]{inputenc}
\usepackage{amsmath}
\usepackage{amsfonts}
\usepackage{amssymb}
\usepackage{braket}
\usepackage{graphicx}
\usepackage{hyperref}
\usepackage{xcolor}
\usepackage{braket}
\usepackage{siunitx}
\newcommand{\mb}{\mathbf}
\newcommand{\bs}{\boldsymbol}
\newcommand{\T}{\text}

\DeclareMathAlphabet{\mathdutchcal}{U}{dutchcal}{m}{n}
\SetMathAlphabet{\mathdutchcal}{bold}{U}{dutchcal}{b}{n}

\begin{document}

\title{Hanbury Brown and Twiss interference of  electrons in free space from independent needle tip sources} 

\newcommand*{\QOQI}{Quantum Optics and Quantum Information, Department of Physics,\\ Friedrich-Alexander-Universität Erlangen-Nürnberg, Staudtstr. 1, 91058 Erlangen, Germany}
\newcommand*{\laser}{Department of Physics, Friedrich-Alexander-Universität Erlangen-Nürnberg, Staudtstr. 1, 91058 Erlangen, Germany}

\author{Anton Classen}
\affiliation{\QOQI}
\affiliation{Health Science Core Facilities,, University of Utah, UT 84112 Salt Lake City, USA}

\author{Raul Corr\^ea}
\affiliation{Departamento de F\'isica, Universidade Federal de Minas Gerais, Belo Horizonte, MG 30123-970, Brazil}

\author{Florian Fleischmann}
\affiliation{\QOQI}

\author{Simon Semmler}
\affiliation{\QOQI}

\author{Marc-Oliver Pleinert}
\affiliation{\QOQI}

\author{Peter Hommelhoff}
\affiliation{\laser}

\author{Joachim von Zanthier}
\affiliation{\QOQI}

\begin{abstract} 
We investigate two-electron interference in free space using two laser-triggered needle tips as independent electron sources, a fermionic realisation of the landmark Hanbury Brown and Twiss interferometer. We calculate the two-electron interference pattern in a quantum path formalism taking into account the fermionic nature and the spin configuration of the electrons. We also estimate the Coulomb repulsion in the setup in a semiclassical approach. We find that antibunching resulting from Pauli's exclusion principle and repulsion stemming from the Coulomb interaction can be clearly distinguished.
\end{abstract}

\maketitle

\section{Introduction}

Since the landmark experiments by Hanbury Brown and Twiss and later by Hong, Ou, and Mandel, it was demonstrated in countless experiments that beyond
single-particle interference also two- and multi-particle interferences are possible even if the particles originate from independent sources and are statistically independent~\cite{Brown:1956,Hanbury-Brown:1956,Fano:1961,Mandel:1982,Hong:1987,Ou:1999,Glauber:2006,Agafonov:2008,Pan:2012,Tichy:2012,Agne:2017,Menssen:2017,Pleinert:2021}.
The essential requirement to record multi-particle interference is the indistinguishability of the particles, i.e., the condition that the detectors upon detection are not able to extract any individual particle information with respect to position, frequency, polarization, and time. 
Yet, in a multi-particle interferometer two fundamentally different kinds of interference exist depending on whether the particles are bosons or fermions. 
This is due to the fact that for bosons the multi-particle wavefunction has to be symmetric, while for fermions it has to be antisymmetric. Correspondingly, for a zero phase delay of all particles in a multi-particle interferometer, bosons display constructive interference, whereas fermions show destructive interference reflecting the Pauli exclusion principle (PEP), which prohibits the simultaneous detection of identical fermions at the same position~\cite{Jeltes:2007}.

In the case of charged fermions, besides displaying anti-correlations due to their fermionic nature, also the Coulomb interaction plays a role. For equally charged fermions, the Coulomb interaction leads to a repulsion of the particles and thus likewise to a reduction of the joint detection probability. Consequently, two electrons -- being fermionic and charged -- in a Hanbury Brown and Twiss (HBT) interferometer will exhibit both fermionic anti-correlation due to PEP as well as Coulomb repulsion due to their charge. Unraveling destructive interference for electrons in a HBT interferometer due to PEP from repulsion due to Coulomb interaction is thus not straightforward and has been proven experimentally to be difficult, even in the most basic case of two-electron interference~\cite{Jeltes:2007,Henny:1999,Oliver:1999,Bocquillon:2013,Kiesel:2002,Lougovski:2011,Baym:2014,Kuwahara:2021,Batelaan:2021,Kuwahara:2021a,Hommelhoff:2006a,Ropers:2007,Barwick:2007,Ehberger:2015,Meier:2018,Keramati:2021,Haindl:2023,Meier:2023}. 

In recent years, multi-photon interferences produced by independent photon emitters, e.g., by spontaneously emitting atoms, have been actively investigated. This is, in particular, due to their potential for super-resolving imaging, e.g., useful for X-ray structure analysis and other fields~\cite{Zou:1991,Lemos:2014,Oppel:2012,Classen:2016,Classen:2017,Schneider:2018,Ho:2021,Richter:2021,Trost:2023}. 
Taking up these ideas and transferring them to multi-electron interference might thus be particular fruitful and open up new paths in electron-based imaging. 
Yet, for the upcoming field of electron correlation spectroscopy, the quantum and classical contributions to multi-electron interference need to be disentangled.

So far, two-electron interference has been demonstrated with electrons propagating in the controlled and screened environment of semiconductor chip devices~\cite{Henny:1999,Oliver:1999,Bocquillon:2013}. 
For electrons propagating in free space, by contrast, there has been controversies on whether true quantum two-electron anti-correlations have been observed~\cite{Kiesel:2002,Lougovski:2011,Baym:2014,Kuwahara:2021,Batelaan:2021,Kuwahara:2021a}. The experiment by Kiesel \textit{et al.}~\cite{Kiesel:2002} investigated far-field correlations between electrons field-emitted by a tungsten needle tip. There were hints for a reduced coincidence rate resulting from PEP, but due to the small source degeneracy, i.e., the number of particles per phase-space-cell volume~\cite{Lougovski:2011}, and the fact that both electrons were created within the same tip, i.e., in close vicinity to each other, a clear verification and distinction of the effect from Coulomb repulsion was difficult~\cite{Lougovski:2011,Baym:2014}. Recently, Kuwahara et al. detected electronic antibunching in a HBT setup of a polarized electron beam~\cite{Kuwahara:2021}. They observed a spin-dependent dip in the time correlation, which they associate to PEP, i.e., the fermionic statistics of the electrons. Due to the spin dependency, Coulomb interaction can be ruled out to contribute significantly to the dip in their setup. However, the shape of the detected temporal correlation carries no spatial information, and experimental asymmetries could as well have explained the result~\cite{Batelaan:2021}. Careful analysis of the experimental parameters in a HBT interferometer for electrons is thus crucial and very important~\cite{Batelaan:2021,Kuwahara:2021a}.

In this paper, we propose to use a different setup to overcome the aforementioned difficulties and observe two-electron quantum interference resulting from PEP in a spatial pattern. 
The idea is to use two or more independent electron emitters like in the original HBT experiment, e.g., two needle tips, where the electrons are emitted by the photoelectric effect.
By using two tips, the period of the HBT oscillations can be adjusted via the distance of the tips in order to unambiguously distinguish a decrease in coincidence detection due to fermionic anti-correlation from the effect of Coulomb repulsion. 
In addition, the Coulomb repulsion of the electrons at the source can be strongly suppressed since the electrons are not created in close vicinity to each other, i.e., within the same tip. 
Laser-triggered tungsten needle tips~\cite{Hommelhoff:2006a,Ropers:2007,Barwick:2007,Ehberger:2015,Meier:2018,Keramati:2021,Haindl:2023,Meier:2023} are ideally suited as electron sources for such experiments, since they can provide high degeneracies and spatial coherence~\cite{Ehberger:2015,Meier:2018} in the emitted wave packets. 
An implementation of the proposed setup with currently available technology can thus be readily achieved.

\section{Independent needle tip sources: Setup and states}

\begin{figure}
  \centering
    \includegraphics[width=0.95\columnwidth]{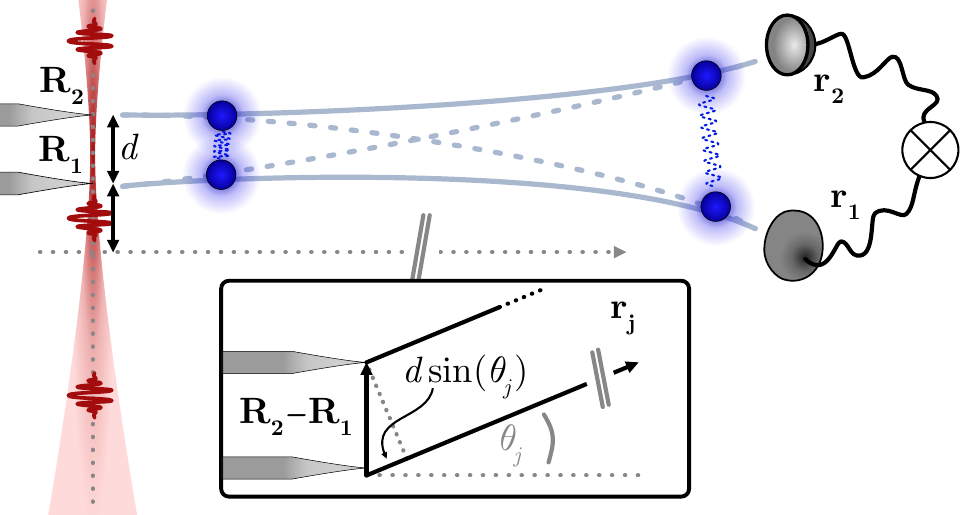}
    \caption{\textbf{Setup of  the system.} Two independent needle tip sources at positions $\mb{R}_1$ and $\mb{R}_2$ with distance $d$ are illuminated by short laser pulses to emit electrons via the photo effect. In the far field, the electrons are coincidentally detected by detectors at positions $\mb{r}_1$ and $\mb{r}_2$. The resulting correlation pattern $G^{(2)}(\mb{r}_1,\mb{r}_2)$ exhibits two-particle interference due to the interference of different two-electron quantum paths. Inset: The path difference for two electrons emitted by different sources but detected at the same detector $j$ is given by $d\sin(\theta_j)$.}\label{fig:setup}
\end{figure}

In what follows, we consider the setup depicted in Fig.~\ref{fig:setup}. 
Two independent electron sources are placed at positions $\mb{R}_l$ ($l=1,2$) with distance $d$, where each source emits electrons in a pulsed manner, e.g., via femtosecond laser pulses. 
The sources are assumed to emit nearly-monochromatic fields at de Broglie wavelength $\lambda_{dB}$. 
This assumption is based on the fact that a pulsed tungsten tip can emit electron wave packets with an energy uncertainty of around $1\%$ \cite{Ehberger:2015}\footnote{In \textit{spatial} second-order correlations, an energy uncertainty of around $1\%$ will merely lead to a slight reduction of the contrast.}. Hereby, we assume that the electrons are accelerated by voltages up to 380V such that the electrons can be considered non-relativistic.
For simplicity, we further assume that the sources are point like, of equal intensity, and of equal spatial mode profile. 
In the proposed setup, the electrons field state is generated by the pulsed laser at both sources simultaneously, and the inability to control the initial phase of the emitted electrons will destroy any coherence between different electron number states \cite{Lougovski:2011}. 
Both tips are thus statistically independent and the resultant initial state is given by the tensor product $\rho=\rho_1\otimes\rho_2$ with
\begin{align}
\rho_l=p_0&|0_l\rangle\langle 0_l|+p_1\sum_{s=1}^2|1_{l,s}\rangle\langle 1_{l,s}|\notag\\&+p_2\sum_{s=1}^2\sum_{s'=1}^2|2_{l,s,s'}\rangle\langle 2_{l,s,s'}| + \ldots \, ,\label{rhol}
\end{align}
for source $l=1,2$.
Here, the ket $|0_l\rangle$ represents no particles being created at source $l$, whereas $|1_{l,s}\rangle$ corresponds to one particle emitted from source $l$ with spin $s$ in an arbitrary nearly-monochromatic mode
\begin{equation}
|1_{l,s}\rangle=\int \!\!d\bs{\mathdutchcal{k}}\ C(\bs{\mathdutchcal{k}})\ a^\dagger_{l,s,\bs{\mathdutchcal{k}}}|0_l\rangle\, .
\label{1e-state}
\end{equation}
Here, $a^\dagger_{l,s,\bs{\mathdutchcal{k}}}$ is the creation operator of an electron at source $l$ with spin $s$ propagating in a plane wave mode with direction $\bs{\mathdutchcal{k}}$ and de Broglie wavelength $\lambda_{dB}=2\pi/|\bs{\mathdutchcal{k}}|$. 
These operators $a_{l,s,\bs{\mathdutchcal{k}}}$ obey the fermionic anti-commutation relations, i.e., $\{a_{l,s,\bs{\mathdutchcal{k}}},a_{l',s',\bs{\mathdutchcal{k}}'}\}=0$ and $\{a_{l,s,\bs{\mathdutchcal{k}}},a^\dagger_{l',s',\bs{\mathdutchcal{k}}'}\}=\delta_{ll'}\delta_{ss'}{\delta(\bs{\mathdutchcal{k}}-\bs{\mathdutchcal{k}}')}$, where $\delta_{ij}$ is the Kronecker delta and $\delta(\bs{\mathdutchcal{k}}-\bs{\mathdutchcal{k}}')$ is the Dirac delta function. 
$C(\bs{\mathdutchcal{k}})$ in Eq.~\eqref{1e-state} are the emission modes in transverse momentum space, and are a complex function of the direction of propagation $\bs{\mathdutchcal{k}}$ being responsible for the envelope in the far-field detection pattern. 
Finally, $|2_{l,s,s'}\rangle$ is a two-electron state in which two particles are created at the same source $l$ in two possibly different spatial modes and different spin states, i.e.,
\begin{equation}
|2_{l,s,s'}\rangle=\int \!\!d\bs{\mathdutchcal{k}}\!\int \!\!d\bs{\mathdutchcal{k}}'\ C_1(\bs{{\mathdutchcal{k}}})C_2(\bs{\mathdutchcal{k}}')\ a^\dagger_{l,s,\bs{\mathdutchcal{k}}}a^\dagger_{l,s',\bs{\mathdutchcal{k}}'}|0_l\rangle.
\label{2e-state}
\end{equation}
In this paper, we consider the probability to emit three electrons by a given source to be negligible and assume that different spin states within a given source are generated with equal probability.

In the far field of the sources at a distance $D$, the electrons are then jointly detected at positions $\mb{r}_j$ ($j=1,2$), e.g., by a delay-line detector with multi-hit capability~\cite{Meier:2023}.
The field operator for electrons of spin $s$ at position $\mb{r}_j$ in the non-relativistic regime (in which the spin of the electron is decoupled from the spatial degrees of freedom) can be written analogously to a photon field~\cite{Skornia:2001,Thiel:2007,Mahrlein:2017} as a sum over the source modes $a_{l,s,\bs{\mathdutchcal{k}}}$, i.e.,
\begin{align}
\Psi_s(\mb{r}_j,t) = \sum_{l=1}^{2}e^{\mathrm{i}\bs{\mathdutchcal{k}}_j\cdot(\mathbf{r}_j-\mathbf{R}_l)-\mathrm{i}\omega(\bs{\mathdutchcal{k}}_j) t}\, a_{l,s,\bs{\mathdutchcal{k}}_j} \, ,\label{farfield-general}
\end{align}
with plane wave modes with a direction towards the $j$th detector, i.e., $\bs{\mathdutchcal{k}} = \bs{\mathdutchcal{k}}_j || \mb{r}_j$.
The vector field operator collecting the different spin states then reads
$\bs{\Psi}(\mb{r}_j,t)=\sum_{s=1}^{2}\Psi_s(\mb{r}_j,t)\bs{\epsilon}_s$, where the unitary spin vectors $\bs{\epsilon}_s$ obey $\bs{\epsilon}_s\cdot\bs{\epsilon}^*_{s'}=\delta_{s,s'}$.

In our setup with nearly-monochromatic fields, the time dependence results in a mere global phase - as does the term $\bs{\mathdutchcal{k}}_j \cdot \mathbf{r}_j \approx 2\pi D/\lambda_{dB}$; thus both can be neglected.
Further noting that we can rewrite $\bs{\mathdutchcal{k}}_j = |\bs{\mathdutchcal{k}}_j| \mb{r}_j / |\mb{r}_j|$, the field operator simplifies to
\begin{align}
\Psi_s(\mb{r}_j) = \sum_{l=1}^{2}\exp\left[-\mathrm{i}\!\left(\frac{2\pi}{\lambda_{dB}}\right)\frac{\mb{r}_j\cdot\mb{R}_l}{|\mb{r}_j|}\right]a_{l,s,\bs{\mathdutchcal{k}}_j} \, .\label{farfield}
\end{align}
Pivotal for two-electron interference is the phase difference between two electrons originating from different sources at $\mb{R}_1$ and $\mb{R}_2$ but being detected at the same detector at $\mb{r}_j$. 
Note that even though the two electrons originate from different tips, they occupy the same modes, i.e., identical $\bs{\mathdutchcal{k}}$ vectors, in the far field~\cite{Mahrlein:2017}.
However, they collect a phase difference given by 
$\delta_j = (2\pi/\lambda_{dB})\mb{r}_j\cdot(\mb{R}_2-\mb{R}_1)/|\mb{r}_j| = 2\pi d\sin(\theta_j)/\lambda_{dB}$, where $d$ is the distance between the two sources and $\theta_j$ the angle between the optical axis and the direction $\mb{r}_j$ of the $j$th detector (see inset in Fig.~\ref{fig:setup}).  
Without loss of generality, we set $\theta_1=0$ ($\delta_1=0$) throughout the paper. $\delta \equiv \delta_2$ will thus be the decisive parameter for the observation of the interference pattern.

\section{Second-order correlations of electrons neglecting charge}

\begin{figure*}
\centering
\includegraphics[width=\textwidth]{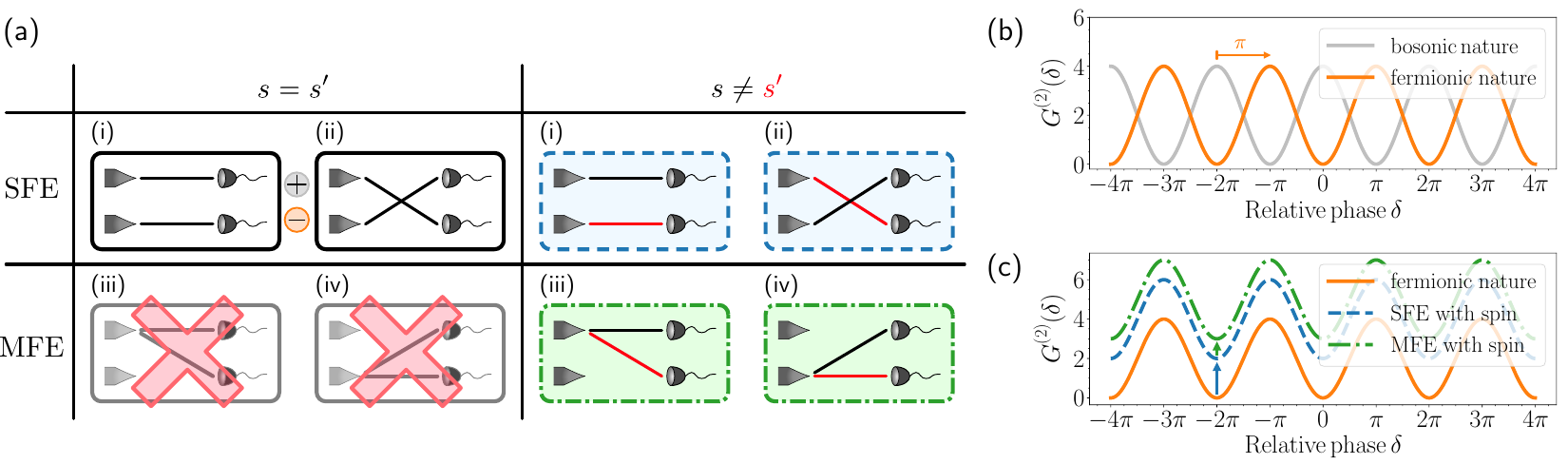}
\caption{\textbf{Quantum paths and resulting second-order spatial correlation functions  $G^{(2)}(\delta)$.} (a) Different possible quantum paths of the electrons - distinguished according to equal spin ($s=s'$) vs. different spin ($s\neq s'$) as well as single-fermion emitters (SFE) emitting a single photon at maximum vs. multi-fermion emitters (MFE), where two electrons can be emitted by a single source. Note that the latter is forbidden for equal spins by the Pauli principle.
(b) For single-particle emitters and neglecting different spins, i.e., concentrating on the two upper left quantum paths of (a), fermionic and bosonic correlations exhibit two-particle interference with visibility one, but with a $\pi$ phase shift showing opposite interfering behaviour. This is due to the different commutativtiy of fermionic and bosonic particles. 
(c) Including different spin settings leads to an offset as shown in blue (dashed) for single-fermion emitters (SFE) due to the additional two quantum paths top right in (a). For multi-fermion emitters (MFE), the two green (dot-dashed) quantum paths displayed bottom right in (a) additionally add to the offset. For the plots, we assume a Poissonian particle distribution.
For a better comparison with the bosonic case, we plot in (b) and (c) $G^{(2)}(\delta)$ with $4p_1^2|C(\bs{\mathdutchcal{k}}_1)|^2|C(\bs{\mathdutchcal{k}}_2)|^2 = 2$.}
\label{fig:Paths_and_Patterns}
\end{figure*}

Neglecting charge, the two-electron correlation function, i.e., the probability to coincidentally detect two electrons at detector positions $\mathbf{r}_1,\mathbf{r}_2$, can be written in terms of the  electron field operators $\mb{\Psi}(\mb{r})$ as
\begin{align}
G^{(2)}(\mb{r}_1,\mb{r}_2):=&\left\langle :\left[\bs{\Psi}^\dagger(\mb{r}_1)\cdot \bs{\Psi}(\mb{r}_1)\right]\ \left[\bs{\Psi}^\dagger(\mb{r}_2)\cdot \bs{\Psi}(\mb{r}_2)\right]:\right\rangle_\rho\notag\\
=&\sum_{s=1}^2\sum_{s'=1}^2 G^{(2)}_{s,s'}(\mb{r}_1,\mb{r}_2)\, ,
\end{align}
where the colons indicate normal ordering of the creation and annihilation operators and the brackets indicate the expectation value to be calculated according to the source state $\rho$. The expression $G^{(2)}(\mb{r}_1,\mb{r}_2)$ can be decomposed into the different contributions of equal and unequal spin polarization via the spin-specific two-electron correlation functions given by
\begin{align}
G^{(2)}_{s,s'}(\mb{r}_1,\mb{r}_2):=\left\langle \Psi_s^\dagger(\mb{r}_1)\Psi_{s'}^\dagger(\mb{r}_2) \Psi_{s'}(\mb{r}_2)\Psi_s(\mb{r}_1)  \right\rangle_\rho \, .\label{gss}
\end{align}
Out of the in total 16 terms, resulting from inserting Eq.~\eqref{farfield} four times into Eq.~\eqref{gss}, only six two-electron contributions survive due to the statistical independence of the two sources yielding~\footnote{Note that any random emission phases cancel in Eq.~\eqref{ggeral} due to the joint appearance of creation and annihilation operator for each source.}
\begin{align}
\langle \Psi_s^\dagger(\mb{r}_1)&\Psi_{s'}^\dagger(\mb{r}_2)\Psi_{s'}(\mb{r}_2)\Psi_s(\mb{r}_1)  \rangle_\rho=\notag\\
&\langle a^\dagger_{1,s,\bs{\mathdutchcal{k}}_1}a^\dagger_{1,s',\bs{\mathdutchcal{k}}_2}a_{1,s',\bs{\mathdutchcal{k}}_2}a_{1,s,\bs{\mathdutchcal{k}}_1}\rangle_\rho\notag\\
&+\langle a^\dagger_{2,s,\bs{\mathdutchcal{k}}_1}a^\dagger_{2,s',\bs{\mathdutchcal{k}}_2}a_{2,s',\bs{\mathdutchcal{k}}_2}a_{2,s,\bs{\mathdutchcal{k}}_1}\rangle_\rho\notag\\
&+\langle a^\dagger_{1,s,\bs{\mathdutchcal{k}}_1}a^\dagger_{2,s',\bs{\mathdutchcal{k}}_2}a_{2,s',\bs{\mathdutchcal{k}}_2}a_{1,s,\bs{\mathdutchcal{k}}_1}\rangle_\rho\notag\\
&+\langle a^\dagger_{2,s,\bs{\mathdutchcal{k}}_1}a^\dagger_{1,s',\bs{\mathdutchcal{k}}_2}a_{1,s',\bs{\mathdutchcal{k}}_2}a_{2,s,\bs{\mathdutchcal{k}}_1}\rangle_\rho\notag\\
&+e^{-i\delta}\langle a^\dagger_{2,s,\bs{\mathdutchcal{k}}_1}a^\dagger_{1,s',\bs{\mathdutchcal{k}}_2}a_{2,s',\bs{\mathdutchcal{k}}_2}a_{1,s,\bs{\mathdutchcal{k}}_1}\rangle_\rho\notag\\
&+e^{i\delta}\langle a^\dagger_{1,s,\bs{\mathdutchcal{k}}_1}a^\dagger_{2,s',\bs{\mathdutchcal{k}}_2}a_{1,s',\bs{\mathdutchcal{k}}_2}a_{2,s,\bs{\mathdutchcal{k}}_1}\rangle_\rho \, .\label{ggeral}
\end{align}
The first four terms in Eq.~\eqref{ggeral} can be identified as the four two-particle paths depicted in Fig.~\ref{fig:Paths_and_Patterns}(a). These correspond to the four different options for the two particles reaching the two detectors: 
both from source 1 (path iii), both from source 2 (path iv), or one from each source, where there are two possibilities (path i) and (path ii). 
Note that the first two terms in Eq.~\eqref{ggeral} are related to the individual statistics of the sources 1 and 2, since each term contains two particle creation from the same source. 
These two-particle terms do not show any oscillating pattern in the far field, and will be responsible for a mere offset in the correlation function $G^{(2)}_{s,s'}(\mb{r}_1,\mb{r}_2)$, when two or more particles from the same source are available, which we call multi-fermion emitters (MFE).

For identical particles (e.g. electrons with equal spin), the two two-particle paths (i) and (ii) [3rd and 4th term in Eq.~\eqref{ggeral}] cannot be distinguished and their coherent addition leads to interference resulting in the two additional terms 5 and 6 in Eq.~\eqref{ggeral}. 
The last four terms in Eq.~\eqref{ggeral} are hence the ones determining the two-particle interference and
due to the phase difference $\delta$ will be responsible for an oscillation pattern in the joint detection probability.
The resulting patterns are shown in Fig.~\ref{fig:Paths_and_Patterns}(b) and display the signature of the type of particles involved. 
For single bosons, the operators of different sources commute resulting in an interference pattern of the kind $|1+\mathrm{e}^{\mathrm{i}\delta}|^2$.
Bosons thus display bunching, i.e., they have a higher chance of being detected at the same point in space and time.
For single fermions, however, the operators anti-commute leading to $|1+\mathrm{e}^{\mathrm{i}(\delta+\pi)}|^2$ with an extra $\pi$ phase shift as shown in Fig.~\ref{fig:Paths_and_Patterns}(b). 
Fermions thus exhibit the opposite behavior and display anti-bunching, i.e., there are never two identical fermions at the same point in space and time, a direct consequence of the Pauli exclusion principle.
Note that the last four terms in Eq.~\eqref{ggeral} are also the ones responsible for the well-known Hong-Ou-Mandel dip of two photons impinging on each side of a beam splitter, where the $\pi$ phase shift does not stem from the particle statistics but is due to the reflection inside the beam splitter~\cite{Hong:1987}.

Having discussed the structure of $G^{(2)}_{s,s'}(\mb{r}_1,\mb{r}_2)$, we now evaluate this expression according to the joint initial state of the two sources $\rho=\rho_1\otimes\rho_2$ given in Eq.~\eqref{rhol}. 
The terms containing less than two particles in total cannot contribute to a two-fold detection event. Thus, we can write
%
%
%
\begin{align}
G^{(2)}_{s,s'}(\delta)= p_1^2 &\sum_{w,w'=1}^2 \langle 1_{w},1_{w'}| \Psi_s^\dagger\Psi_{s'}^\dagger\Psi_{s'}\Psi_s  |1_{w},1_{w'}\rangle\label{gss-solved} \\
+ 2p_0p_2&\sum_{w,w'=1}^2\langle 2_{w,w'},0| \Psi_s^\dagger\Psi_{s'}^\dagger\Psi_{s'}\Psi_s  |2_{w,w'},0\rangle\,, \notag
\end{align}
where we dropped the subscript $l$ of the source in the two-mode states, because it is implicit in the order of appearance, i.e., $|1_{1,s},1_{2,s'}\rangle \equiv |1_s,1_{s'}\rangle$: and further used the notation $\Psi_s^\dagger\Psi_{s'}^\dagger\Psi_{s'}\Psi_s  \equiv \Psi_s^\dagger(\mb{r}_1)\Psi_{s'}^\dagger(\mb{r}_2)\Psi_{s'}(\mb{r}_2)\Psi_s(\mb{r}_1) $. 
Note that for our choice of detector positions ($\mb{r}_1$ fixed at $\theta_1=0$), the final pattern only depends on $\delta_2=\delta$.
Note that in Eq.~\eqref{gss-solved}, we have considered that $p_0\gg p_1\gg p_2$, so that the terms of order $p_1p_2$ and $p_2^2$ can be neglected. 
There are thus two main contributions in Eq.~\eqref{gss-solved}: One electron from each source $G^{(2)}_{s,s',p_1^2}(\delta)$ [first line in Eq.~\eqref{gss-solved}] and two electrons coming from the same source $G^{(2)}_{s,s',p_0p_2}(\delta)$ [second line in Eq.~\eqref{gss-solved}].
Note that typically $p_1^2$ and $p_0p_2$ are of the same order of magnitude (e.g., for a thermal source or a source with a Poissonian distribution).
We have also made use of the fact that the spatial modes emitted are independent of spin and equal for each source, hence the factor of $2$ in front of the second term.

For the state $|1_{w},1_{w'}\rangle$, $G^{(2)}_{s,s',p_1^2}(\delta)$ will only be nonzero if $w=s$ and $w'=s'$ or vice versa leading to the two paths (i) and (ii) in Fig.~\ref{fig:Paths_and_Patterns}(a). 
The state has at most one electron coming from each source, hence the first two terms in Eq.~\eqref{ggeral} will be zero. Further, the last two terms in Eq.~\eqref{ggeral} will only be nonzero for $s=s'$, fulfilling the expectation that no interfering oscillation will appear if the detected particles have orthogonal spins. Separating the terms with equal and unequal spin, we thus obtain
\begin{align}
G^{(2)}_{s=s',p_1^2}(\delta)&=4p_1^2 |C(\bs{\mathdutchcal{k}}_1)|^2|C(\bs{\mathdutchcal{k}}_2)|^2\big[1-\cos(\delta)\big],\label{interference}\\
G^{(2)}_{s\neq s',p_1^2}(\delta)&=4p_1^2 |C(\bs{\mathdutchcal{k}}_1)|^2|C(\bs{\mathdutchcal{k}}_2)|^2.\label{interference-neq}
\end{align}

For the state $|2_{w,w'},0\rangle$, in which both electrons originate from the same source, the situation is reversed.
The four last terms in Eq.~\eqref{ggeral} are zero, and only the first two terms survive, resulting in
\begin{widetext}
\begin{align}
G^{(2)}_{s,s',p_0p_2}(\delta)=4p_0p_2\sum_{w,w'=1}^2&\Big\{\delta_{w,s}\delta_{w',s'}|C_A(\bs{\mathdutchcal{k}}_1)|^2|C_B(\bs{\mathdutchcal{k}}_2)|^2+\delta_{w,s'}\delta_{w',s}|C_A(\bs{\mathdutchcal{k}}_2)|^2|C_B(\bs{\mathdutchcal{k}}_1)|^2\notag\\
&-2\delta_{w,s}\delta_{w,s'}\delta_{w',s}\delta_{w',s'}\T{Re}\big[C_A(\bs{\mathdutchcal{k}}_1)C_A^*(\bs{\mathdutchcal{k}}_2)C_B(\bs{\mathdutchcal{k}}_2)C_B^*(\bs{\mathdutchcal{k}}_1)\big]\Big\}.
\label{g2-2el}
\end{align}
\end{widetext}
For unequal spins of the two electrons $s\neq s'$, the first and second term correspond to the green (dot-dashed border) paths (iii) and (iv) in Fig.~\ref{fig:Paths_and_Patterns}(a), while the third term is zero.
Assuming that all emission modes of the two sources are identical and independent of the number of electrons emitted, i.e.,  $C_A(\bs{\mathdutchcal{k}})=C_B(\bs{\mathdutchcal{k}})=C(\bs{\mathdutchcal{k}})$, then the last term in the above equation will cancel the first two. Thus, there is no contribution from two electrons with equal spin emitted by the same source reflecting that due to the PEP the source cannot emit two identical electrons, i.e., two monochromatic electrons with the same wavelength, in the same spatial mode, and with the same spin.
In total, the contribution by two electrons from the same source thus reads
\begin{align}
G^{(2)}_{s=s',p_0p_2}(\delta)&=0\\
G^{(2)}_{s\neq s',p_0p_2}(\delta)&=4p_0p_2 |C(\bs{\mathdutchcal{k}}_1)|^2|C(\bs{\mathdutchcal{k}}_2)|^2 \,.
\end{align}

Summing the contributions of one- and two-electron emissions per tip, we obtain
\begin{align}
G^{(2)}_{s=s'}(\delta)&=4p_1^2 |C(\bs{\mathdutchcal{k}}_1)|^2|C(\bs{\mathdutchcal{k}}_2)|^2\big[1-\cos(\delta)\big],\\
G^{(2)}_{s\neq s'}(\delta)&=4(p_1^2+p_0p_2) |C(\bs{\mathdutchcal{k}}_1)|^2|C(\bs{\mathdutchcal{k}}_2)|^2.
\end{align}
Allowing for arbitrary spin configurations, the far-field second-order correlation function, in the nearly-monochromatic regime, thus reads
\begin{align}
G^{(2)}(\delta)&=4p_1^2|C(\bs{\mathdutchcal{k}}_1)|^2|C(\bs{\mathdutchcal{k}}_2)|^2\left[2+\frac{p_0p_2}{p_1^2}-\cos(\delta)\right],\label{final-pattern}
\end{align}
for which the visibility is given by $\mathcal{V}=1/(2+p_0p_2/p_1^2)$.

In Fig.~\ref{fig:Paths_and_Patterns}(c), the resulting second-order correlation function $G^{(2)}(\delta)$ is shown for three different fermionic sources: In orange (solid) for single-fermion emitters with polarized spins [Eq.~\eqref{interference}], in blue (dashed) for single-fermion emitters with unpolarized spins [Eq.~\eqref{final-pattern} with $p_2=0$], and in green (dot-dashed) for multi-fermion emitters with unpolarized spins [Eq.~\eqref{final-pattern} with Poissonian statistics].
For ideal single-fermion emitters ($p_2 = 0$) and perfect polarization of the spins ($s=s'$), the interference pattern exhibits a contrast of $100\%$ (orange solid curve), just like in the photonic case \cite{Mandel:1983,Skornia:2001}, but with complementary fringes due to the fermionic anti-commutation nature. 
For unpolarized spins, but still single-fermion emitters, the contrast reduces to $50\%$ because of the constant offset due to the additional unequal spin configurations (blue dashed curve). 
For multi-fermion emitters with $p_2 > 0$, the contrast further reduces, e.g., for a Poissonian statistics to $40\%$.
The contrast of the detected two-electron correlation function can thus also be used to estimate the electron statistics of the laser-driven tungsten tip sources.

\section{Classical estimation of the Coulomb repulsion}

\begin{figure}
\centering
\includegraphics[width=0.85\columnwidth]{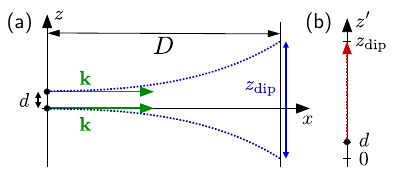}
\caption{\textbf{Scheme for the classical estimation of the Coulomb repulsion.} (a) Two electrons are emitted with parallel momentum $\mathbf{k}$ towards the detection screen. Due to the Coulomb force, the two electrons repel each other (blue dotted trajectories) leading to an increased separation of the electrons and an estimated dip width $z_{\mathrm{dip}}$. (b) In the centre-of-mass frame, the problem can be reduced to the one-dimensional movement of an electron in a Coulomb potential.}
\label{ClassicalElectron}
\end{figure}

So far, we neglected the charge of the electrons.
The resulting Coulomb interaction leads to a repulsion between the electrons and thus to a (Coulomb) dip in the second-order correlation function $G^{(2)}(\delta)$
- as does the Pauli exclusion principle due to the fermionic nature. 
We will show that in a single-tip setup these two effects can be easily confused, while in the proposed two-tip setup they can be clearly distinguished.

To estimate the size of the Coulomb dip, we consider the following classical model: Two electrons separated by the tip distance $d$ are being ejected with momentum $\hbar\mathbf{k}$ towards the detector screen located at a distance $D$ as illustrated in Fig.~ \ref{ClassicalElectron}(a). 
The repulsive Coulomb force between the two electrons is given by
\begin{equation}
    \mathbf{F}_{12}(\mathbf{r_1},\mathbf{r_2})=\frac{\mathrm{e}^2}{4 \pi \epsilon_0} \frac{\mathbf{r_1}-\mathbf{r_2}}{\left|\mathbf{r_1}-\mathbf{r_2}\right|^3}.
\end{equation}
where $e$ is the elementary charge, $\epsilon_0$ the vacuum permittivity, and $\mathbf{r_1}$ and $\mathbf{r_2}$ denote the positions of the two electrons along their trajectory. 
Since the force only depends on the relative distance, we can consider it equivalently in the one-dimensional relative frame [see Fig.~\ref{ClassicalElectron}(b)], where the equation of motion reads
\begin{equation}
    \ddot z = \frac{e^2}{4 \pi \epsilon_0 m} \frac{1}{z^2}\, ,
    \label{Diff}
\end{equation}
which can be solved numerically, e.g., by using a Verlet approach~\cite{Verlet:1967}. 
As it turns out, the acceleration process of the electron is very quick compared to the timescale given by the time of flight of the centre of mass of the electrons towards the detector. Thus, we can use the asymptotic end velocity of the electrons $v_\mathrm{rel,end}$ to compute the separation of the two electrons after the time of flight $t_\mathrm{f}$ leading to
\begin{equation}
    z_\text{dip}=z(t_\mathrm{f})=\sqrt{\frac{m_e e^2}{\hbar^2 \pi \epsilon_0}} \frac{D}{\sqrt{d}k},
    \label{WIDTHEstimate}
\end{equation}
where we write $t_\mathrm{f}=D/v_\mathrm{cms}$ with $v_\mathrm{cms}=\hbar k/m_e$ the velocity of the centre of mass towards the detector screen.  
For a normal distribution of the initial momentum around $k$, the resulting separations will be normally distributed around $z_\text{dip}$.
We thus model the Coulomb dip as an inverted normal distribution with dip width (FWHM) $z_\text{dip}$.

\begin{figure}
  \centering
    \includegraphics[width=0.99\columnwidth]{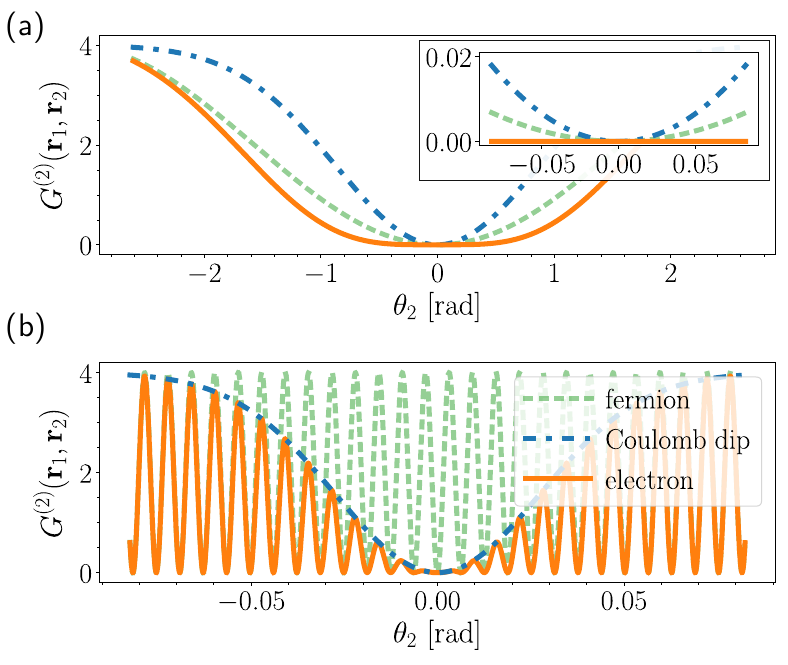}
    \caption{\textbf{Two-electron spatial correlation pattern including Coulomb repulsion,} plotted for perfect spin polarization ($s=s'$), fixed magnitude of the wave vector $k=\SI{1e11}{\per\metre}$ ($\lambda_{dB}=\SI{6.28e-11}{\metre}$), a fixed distance between the needle tip electron sources and the detection screen $D=1  \SI{}{\metre}$, but varying distance between the tips $d$. (a) $d=\SI{0.01}{\nano\metre}$ mimicking a single tip setup; here the oscillation due to Coulomb repulsion and fermionic anti-correlation is of the same order of magnitude.
    (b) $d=\SI{10}{\nano\metre}$ for a realistic two tip setup; here the fermionic oscillations can be clearly differentiated from the Coulomb dip. As in Fig.~\ref{fig:Paths_and_Patterns}, we plot $G^{(2)}(\delta)$ with $4p_1^2|C(\bs{\mathdutchcal{k}}_1)|^2|C(\bs{\mathdutchcal{k}}_2)|^2 = 2$.} \label{fig:CoulombVSPauli}
\end{figure}

The second-order correlation function $G^{(2)}(\delta)$  is then described by the modulation of the fermionic interference pattern with an envelope due to the Coulomb dip. 
A central dip in such a pattern might then be due to the fermionic nature or the Coulomb repulsion of the two electrons. 
We show the expected patterns in Fig.~\ref{fig:CoulombVSPauli}, where we focus on parallel spin configurations ($s=s'$) for clarity. 
In~(a), the tip distance is chosen as $d=\SI{0.01}{\nano\metre}$ mimicking a single-tip setup.
In~(b), we display the same pattern in a two-tip experiment with $d=\SI{10}{\nano\metre}$.

In a single-tip experiment, the wavelength of the interference pattern is of the same magnitude as the width of the Coulomb dip. Distinguishing the two effects is thus rather difficult and the central dip can be attributed to both Coulomb repulsion and Pauli exclusion principle alike. 
In a two-tip setup, however, as proposed in this paper, the influence of Coulomb repulsion is strongly mitigated, due to a larger spatial separation of the sources and thus a higher frequency of the interference pattern. Many interference fringes due to the fermionic anti-correlations resulting from the PEP appear within the size of the Coulomb dip and thus a clear distinction between both effects is possible.

The number of fringes within the Coulomb dip can be calculated via the ratio of the dip width $z_\text{dip}$ and the spatial wavelength $\Lambda_\mathrm{sp}=2\pi D/(k d)$ of the fermionic interference pattern, i.e.,
\begin{equation} \label{eq:number-of-fringes}
    N=\frac{z_\text{dip}}{\Lambda_\mathrm{sp}}=\sqrt{\frac{e^2}{\pi \epsilon_0}\frac{m_e}{h^2}} \cdot \sqrt{d} \propto \sqrt{d} \, .
\end{equation}
As can be seen from Eq.~\eqref{eq:number-of-fringes}, the number of fringes solely depends on the separation of the two tips.
For a single-tip setup with $d \ll \SI{1}{\nm}$, the number of oscillations within the Coulomb dip goes to zero regardless of the other parameters, and the distinction between quantum effects and Coulomb repulsion will be very hard.\\
\vspace{-0.1cm}
\section{Conclusion}

In summary, we showed that in two-electron interference experiments in free space the use of two independent laser-driven needle tip electron sources allows for a clear separation of antibunching due to Pauli's exclusion principle for fermions from Coulomb repulsion.
We first calculated the second-order correlation function for two electrons $G^{(2)}(\delta)$ in a quantum path formalism taking into account the fermionic nature of the electrons and their spin, while neglecting their charge. The latter was included later on in a classical estimation of the Coulomb repulsion.
We demonstrated that the two-tip setup is a highly advantageous configuration, since the spatial two-electron interference pattern arising due to their fermionic nature shows a faster oscillation and thus can be clearly distinguished from the Coulomb repulsion dip, in contrast to a single tip setup. 
In the future, a quantum-mechanical treatment of the Coulomb interaction in a fermionic Hanbury Brown and Twiss interferometer could be addressed treating Coulomb repulsion and fermionic antibunching on an equal basis.
Our results might lead to new approaches for electron imaging, where incoherent electron sources themselves or incoherently scattered electrons in combination with correlation measurements would reveal spatial information.

\paragraph*{Acknowledgments.}
The authors thank Stefan Meier and Jonas Heimerl for fruitful discussions. This work was funded by the Deutsche Forschungsgemeinschaft (DFG, German Research Foundation) – Project-ID 429529648 – TRR 306 QuCoLiMa (“Quantum Cooperativity of Light and Matter’’).
R.C. was funded by the Brazilian agency CAPES, Programa Doutorado Sanduíche no Exterior (PDSE), 
process number 88881.134204/2016-01.
A.C., S.S., M.-O.P., and J.v.Z. acknowledge funding by the Erlangen Graduate School in Advanced Optical Technologies (SAOT) by the Bavarian State Ministry for Science and Art.

\bibliography{ref}

\begin{thebibliography}{47}%
\makeatletter
\providecommand \@ifxundefined [1]{%
 \@ifx{#1\undefined}
}%
\providecommand \@ifnum [1]{%
 \ifnum #1\expandafter \@firstoftwo
 \else \expandafter \@secondoftwo
 \fi
}%
\providecommand \@ifx [1]{%
 \ifx #1\expandafter \@firstoftwo
 \else \expandafter \@secondoftwo
 \fi
}%
\providecommand \natexlab [1]{#1}%
\providecommand \enquote  [1]{``#1''}%
\providecommand \bibnamefont  [1]{#1}%
\providecommand \bibfnamefont [1]{#1}%
\providecommand \citenamefont [1]{#1}%
\providecommand \href@noop [0]{\@secondoftwo}%
\providecommand \href [0]{\begingroup \@sanitize@url \@href}%
\providecommand \@href[1]{\@@startlink{#1}\@@href}%
\providecommand \@@href[1]{\endgroup#1\@@endlink}%
\providecommand \@sanitize@url [0]{\catcode `\\12\catcode `\$12\catcode
  `\&12\catcode `\#12\catcode `\^12\catcode `\_12\catcode `\%12\relax}%
\providecommand \@@startlink[1]{}%
\providecommand \@@endlink[0]{}%
\providecommand \url  [0]{\begingroup\@sanitize@url \@url }%
\providecommand \@url [1]{\endgroup\@href {#1}{\urlprefix }}%
\providecommand \urlprefix  [0]{URL }%
\providecommand \Eprint [0]{\href }%
\providecommand \doibase [0]{https://doi.org/}%
\providecommand \selectlanguage [0]{\@gobble}%
\providecommand \bibinfo  [0]{\@secondoftwo}%
\providecommand \bibfield  [0]{\@secondoftwo}%
\providecommand \translation [1]{[#1]}%
\providecommand \BibitemOpen [0]{}%
\providecommand \bibitemStop [0]{}%
\providecommand \bibitemNoStop [0]{.\EOS\space}%
\providecommand \EOS [0]{\spacefactor3000\relax}%
\providecommand \BibitemShut  [1]{\csname bibitem#1\endcsname}%
\let\auto@bib@innerbib\@empty
\bibitem [{\citenamefont {Hanbury~Brown}\ and\ \citenamefont
  {Twiss}(1956{\natexlab{a}})}]{Brown:1956}%
  \BibitemOpen
  \bibfield  {author} {\bibinfo {author} {\bibfnamefont {R.}~\bibnamefont
  {Hanbury~Brown}}\ and\ \bibinfo {author} {\bibfnamefont {R.~Q.}\ \bibnamefont
  {Twiss}},\ }\bibfield  {title} {\bibinfo {title} {Correlation between photons
  in two coherent beams of light},\ }\href@noop {} {\bibfield  {journal}
  {\bibinfo  {journal} {Nature}\ }\textbf {\bibinfo {volume} {177}},\ \bibinfo
  {pages} {27} (\bibinfo {year} {1956}{\natexlab{a}})}\BibitemShut {NoStop}%
\bibitem [{\citenamefont {Hanbury~Brown}\ and\ \citenamefont
  {Twiss}(1956{\natexlab{b}})}]{Hanbury-Brown:1956}%
  \BibitemOpen
  \bibfield  {author} {\bibinfo {author} {\bibfnamefont {R.}~\bibnamefont
  {Hanbury~Brown}}\ and\ \bibinfo {author} {\bibfnamefont {R.~Q.}\ \bibnamefont
  {Twiss}},\ }\bibfield  {title} {\bibinfo {title} {A test of a new type of
  stellar interferometer on sirius},\ }\href@noop {} {\bibfield  {journal}
  {\bibinfo  {journal} {Nature}\ }\textbf {\bibinfo {volume} {178}},\ \bibinfo
  {pages} {1046} (\bibinfo {year} {1956}{\natexlab{b}})}\BibitemShut {NoStop}%
\bibitem [{\citenamefont {Fano}(1961)}]{Fano:1961}%
  \BibitemOpen
  \bibfield  {author} {\bibinfo {author} {\bibfnamefont {U.}~\bibnamefont
  {Fano}},\ }\bibfield  {title} {\bibinfo {title} {{Quantum Theory of
  Interference Effects in the Mixing of Light from Phase-Independent
  Sources}},\ }\href@noop {} {\bibfield  {journal} {\bibinfo  {journal}
  {American Journal of Physics}\ }\textbf {\bibinfo {volume} {29}},\ \bibinfo
  {pages} {539} (\bibinfo {year} {1961})}\BibitemShut {NoStop}%
\bibitem [{\citenamefont {Mandel}(1982)}]{Mandel:1982}%
  \BibitemOpen
  \bibfield  {author} {\bibinfo {author} {\bibfnamefont {L.}~\bibnamefont
  {Mandel}},\ }\bibfield  {title} {\bibinfo {title} {Squeezed states and
  sub-poissonian photon statistics},\ }\href@noop {} {\bibfield  {journal}
  {\bibinfo  {journal} {Phys. Rev. Lett.}\ }\textbf {\bibinfo {volume} {49}},\
  \bibinfo {pages} {136} (\bibinfo {year} {1982})}\BibitemShut {NoStop}%
\bibitem [{\citenamefont {Hong}\ \emph {et~al.}(1987)\citenamefont {Hong},
  \citenamefont {Ou},\ and\ \citenamefont {Mandel}}]{Hong:1987}%
  \BibitemOpen
  \bibfield  {author} {\bibinfo {author} {\bibfnamefont {C.~K.}\ \bibnamefont
  {Hong}}, \bibinfo {author} {\bibfnamefont {Z.~Y.}\ \bibnamefont {Ou}},\ and\
  \bibinfo {author} {\bibfnamefont {L.}~\bibnamefont {Mandel}},\ }\bibfield
  {title} {\bibinfo {title} {Measurement of subpicosecond time intervals
  between two photons by interference},\ }\href@noop {} {\bibfield  {journal}
  {\bibinfo  {journal} {Phys. Rev. Lett.}\ }\textbf {\bibinfo {volume} {59}},\
  \bibinfo {pages} {2044} (\bibinfo {year} {1987})}\BibitemShut {NoStop}%
\bibitem [{\citenamefont {Ou}\ \emph {et~al.}(1999)\citenamefont {Ou},
  \citenamefont {Rhee},\ and\ \citenamefont {Wang}}]{Ou:1999}%
  \BibitemOpen
  \bibfield  {author} {\bibinfo {author} {\bibfnamefont {Z.~Y.}\ \bibnamefont
  {Ou}}, \bibinfo {author} {\bibfnamefont {J.-K.}\ \bibnamefont {Rhee}},\ and\
  \bibinfo {author} {\bibfnamefont {L.~J.}\ \bibnamefont {Wang}},\ }\bibfield
  {title} {\bibinfo {title} {Observation of four-photon interference with a
  beam splitter by pulsed parametric down-conversion},\ }\href@noop {}
  {\bibfield  {journal} {\bibinfo  {journal} {Phys. Rev. Lett.}\ }\textbf
  {\bibinfo {volume} {83}},\ \bibinfo {pages} {959} (\bibinfo {year}
  {1999})}\BibitemShut {NoStop}%
\bibitem [{\citenamefont {Glauber}(2006)}]{Glauber:2006}%
  \BibitemOpen
  \bibfield  {author} {\bibinfo {author} {\bibfnamefont {R.~J.}\ \bibnamefont
  {Glauber}},\ }\bibfield  {title} {\bibinfo {title} {Nobel lecture: One
  hundred years of light quanta},\ }\href@noop {} {\bibfield  {journal}
  {\bibinfo  {journal} {Rev. Mod. Phys.}\ }\textbf {\bibinfo {volume} {78}},\
  \bibinfo {pages} {1267} (\bibinfo {year} {2006})}\BibitemShut {NoStop}%
\bibitem [{\citenamefont {Agafonov}\ \emph {et~al.}(2008)\citenamefont
  {Agafonov}, \citenamefont {Chekhova}, \citenamefont {Iskhakov},\ and\
  \citenamefont {Penin}}]{Agafonov:2008}%
  \BibitemOpen
  \bibfield  {author} {\bibinfo {author} {\bibfnamefont {I.~N.}\ \bibnamefont
  {Agafonov}}, \bibinfo {author} {\bibfnamefont {M.~V.}\ \bibnamefont
  {Chekhova}}, \bibinfo {author} {\bibfnamefont {T.~S.}\ \bibnamefont
  {Iskhakov}},\ and\ \bibinfo {author} {\bibfnamefont {A.~N.}\ \bibnamefont
  {Penin}},\ }\bibfield  {title} {\bibinfo {title} {High-visibility multiphoton
  interference of {H}anbury {B}rown--{T}wiss type for classical light},\
  }\href@noop {} {\bibfield  {journal} {\bibinfo  {journal} {Phys. Rev. A}\
  }\textbf {\bibinfo {volume} {77}},\ \bibinfo {pages} {053801} (\bibinfo
  {year} {2008})}\BibitemShut {NoStop}%
\bibitem [{\citenamefont {Pan}\ \emph {et~al.}(2012)\citenamefont {Pan},
  \citenamefont {Chen}, \citenamefont {Lu}, \citenamefont {Weinfurter},
  \citenamefont {Zeilinger},\ and\ \citenamefont {\ifmmode~\dot{Z}\else
  \.{Z}\fi{}ukowski}}]{Pan:2012}%
  \BibitemOpen
  \bibfield  {author} {\bibinfo {author} {\bibfnamefont {J.-W.}\ \bibnamefont
  {Pan}}, \bibinfo {author} {\bibfnamefont {Z.-B.}\ \bibnamefont {Chen}},
  \bibinfo {author} {\bibfnamefont {C.-Y.}\ \bibnamefont {Lu}}, \bibinfo
  {author} {\bibfnamefont {H.}~\bibnamefont {Weinfurter}}, \bibinfo {author}
  {\bibfnamefont {A.}~\bibnamefont {Zeilinger}},\ and\ \bibinfo {author}
  {\bibfnamefont {M.}~\bibnamefont {\ifmmode~\dot{Z}\else \.{Z}\fi{}ukowski}},\
  }\bibfield  {title} {\bibinfo {title} {Multiphoton entanglement and
  interferometry},\ }\href@noop {} {\bibfield  {journal} {\bibinfo  {journal}
  {Rev. Mod. Phys.}\ }\textbf {\bibinfo {volume} {84}},\ \bibinfo {pages} {777}
  (\bibinfo {year} {2012})}\BibitemShut {NoStop}%
\bibitem [{\citenamefont {Tichy}\ \emph {et~al.}(2012)\citenamefont {Tichy},
  \citenamefont {Tiersch}, \citenamefont {Mintert},\ and\ \citenamefont
  {Buchleitner}}]{Tichy:2012}%
  \BibitemOpen
  \bibfield  {author} {\bibinfo {author} {\bibfnamefont {M.~C.}\ \bibnamefont
  {Tichy}}, \bibinfo {author} {\bibfnamefont {M.}~\bibnamefont {Tiersch}},
  \bibinfo {author} {\bibfnamefont {F.}~\bibnamefont {Mintert}},\ and\ \bibinfo
  {author} {\bibfnamefont {A.}~\bibnamefont {Buchleitner}},\ }\bibfield
  {title} {\bibinfo {title} {Many-particle interference beyond many-boson and
  many-fermion statistics},\ }\href@noop {} {\bibfield  {journal} {\bibinfo
  {journal} {New Journal of Physics}\ }\textbf {\bibinfo {volume} {14}},\
  \bibinfo {pages} {093015} (\bibinfo {year} {2012})}\BibitemShut {NoStop}%
\bibitem [{\citenamefont {Agne}\ \emph {et~al.}(2017)\citenamefont {Agne},
  \citenamefont {Kauten}, \citenamefont {Jin}, \citenamefont {Meyer-Scott},
  \citenamefont {Salvail}, \citenamefont {Hamel}, \citenamefont {Resch},
  \citenamefont {Weihs},\ and\ \citenamefont {Jennewein}}]{Agne:2017}%
  \BibitemOpen
  \bibfield  {author} {\bibinfo {author} {\bibfnamefont {S.}~\bibnamefont
  {Agne}}, \bibinfo {author} {\bibfnamefont {T.}~\bibnamefont {Kauten}},
  \bibinfo {author} {\bibfnamefont {J.}~\bibnamefont {Jin}}, \bibinfo {author}
  {\bibfnamefont {E.}~\bibnamefont {Meyer-Scott}}, \bibinfo {author}
  {\bibfnamefont {J.~Z.}\ \bibnamefont {Salvail}}, \bibinfo {author}
  {\bibfnamefont {D.~R.}\ \bibnamefont {Hamel}}, \bibinfo {author}
  {\bibfnamefont {K.~J.}\ \bibnamefont {Resch}}, \bibinfo {author}
  {\bibfnamefont {G.}~\bibnamefont {Weihs}},\ and\ \bibinfo {author}
  {\bibfnamefont {T.}~\bibnamefont {Jennewein}},\ }\bibfield  {title} {\bibinfo
  {title} {Observation of genuine three-photon interference},\ }\href@noop {}
  {\bibfield  {journal} {\bibinfo  {journal} {Phys. Rev. Lett.}\ }\textbf
  {\bibinfo {volume} {118}},\ \bibinfo {pages} {153602} (\bibinfo {year}
  {2017})}\BibitemShut {NoStop}%
\bibitem [{\citenamefont {Menssen}\ \emph {et~al.}(2017)\citenamefont
  {Menssen}, \citenamefont {Jones}, \citenamefont {Metcalf}, \citenamefont
  {Tichy}, \citenamefont {Barz}, \citenamefont {Kolthammer},\ and\
  \citenamefont {Walmsley}}]{Menssen:2017}%
  \BibitemOpen
  \bibfield  {author} {\bibinfo {author} {\bibfnamefont {A.~J.}\ \bibnamefont
  {Menssen}}, \bibinfo {author} {\bibfnamefont {A.~E.}\ \bibnamefont {Jones}},
  \bibinfo {author} {\bibfnamefont {B.~J.}\ \bibnamefont {Metcalf}}, \bibinfo
  {author} {\bibfnamefont {M.~C.}\ \bibnamefont {Tichy}}, \bibinfo {author}
  {\bibfnamefont {S.}~\bibnamefont {Barz}}, \bibinfo {author} {\bibfnamefont
  {W.~S.}\ \bibnamefont {Kolthammer}},\ and\ \bibinfo {author} {\bibfnamefont
  {I.~A.}\ \bibnamefont {Walmsley}},\ }\bibfield  {title} {\bibinfo {title}
  {Distinguishability and many-particle interference},\ }\href@noop {}
  {\bibfield  {journal} {\bibinfo  {journal} {Phys. Rev. Lett.}\ }\textbf
  {\bibinfo {volume} {118}},\ \bibinfo {pages} {153603} (\bibinfo {year}
  {2017})}\BibitemShut {NoStop}%
\bibitem [{\citenamefont {Pleinert}\ \emph {et~al.}(2021)\citenamefont
  {Pleinert}, \citenamefont {Rueda}, \citenamefont {Lutz},\ and\ \citenamefont
  {von Zanthier}}]{Pleinert:2021}%
  \BibitemOpen
  \bibfield  {author} {\bibinfo {author} {\bibfnamefont {M.-O.}\ \bibnamefont
  {Pleinert}}, \bibinfo {author} {\bibfnamefont {A.}~\bibnamefont {Rueda}},
  \bibinfo {author} {\bibfnamefont {E.}~\bibnamefont {Lutz}},\ and\ \bibinfo
  {author} {\bibfnamefont {J.}~\bibnamefont {von Zanthier}},\ }\bibfield
  {title} {\bibinfo {title} {Testing higher-order quantum interference with
  many-particle states},\ }\href@noop {} {\bibfield  {journal} {\bibinfo
  {journal} {Phys. Rev. Lett.}\ }\textbf {\bibinfo {volume} {126}},\ \bibinfo
  {pages} {190401} (\bibinfo {year} {2021})}\BibitemShut {NoStop}%
\bibitem [{\citenamefont {Jeltes}\ \emph {et~al.}(2007)\citenamefont {Jeltes},
  \citenamefont {McNamara}, \citenamefont {Hogervorst}, \citenamefont {Vassen},
  \citenamefont {Krachmalnicoff}, \citenamefont {Schellekens}, \citenamefont
  {Perrin}, \citenamefont {Chang}, \citenamefont {Boiron}, \citenamefont
  {Aspect},\ and\ \citenamefont {Westbrook}}]{Jeltes:2007}%
  \BibitemOpen
  \bibfield  {author} {\bibinfo {author} {\bibfnamefont {T.}~\bibnamefont
  {Jeltes}}, \bibinfo {author} {\bibfnamefont {J.~M.}\ \bibnamefont
  {McNamara}}, \bibinfo {author} {\bibfnamefont {W.}~\bibnamefont
  {Hogervorst}}, \bibinfo {author} {\bibfnamefont {W.}~\bibnamefont {Vassen}},
  \bibinfo {author} {\bibfnamefont {V.}~\bibnamefont {Krachmalnicoff}},
  \bibinfo {author} {\bibfnamefont {M.}~\bibnamefont {Schellekens}}, \bibinfo
  {author} {\bibfnamefont {A.}~\bibnamefont {Perrin}}, \bibinfo {author}
  {\bibfnamefont {H.}~\bibnamefont {Chang}}, \bibinfo {author} {\bibfnamefont
  {D.}~\bibnamefont {Boiron}}, \bibinfo {author} {\bibfnamefont
  {A.}~\bibnamefont {Aspect}},\ and\ \bibinfo {author} {\bibfnamefont {C.~I.}\
  \bibnamefont {Westbrook}},\ }\bibfield  {title} {\bibinfo {title} {Comparison
  of the {H}anbury {B}rown--{T}wiss effect for bosons and fermions},\
  }\href@noop {} {\bibfield  {journal} {\bibinfo  {journal} {Nature}\ }\textbf
  {\bibinfo {volume} {445}},\ \bibinfo {pages} {402} (\bibinfo {year}
  {2007})}\BibitemShut {NoStop}%
\bibitem [{\citenamefont {Henny}\ \emph {et~al.}(1999)\citenamefont {Henny},
  \citenamefont {Oberholzer}, \citenamefont {Strunk}, \citenamefont {Heinzel},
  \citenamefont {Ensslin}, \citenamefont {Holland},\ and\ \citenamefont
  {Sch{\"o}nenberger}}]{Henny:1999}%
  \BibitemOpen
  \bibfield  {author} {\bibinfo {author} {\bibfnamefont {M.}~\bibnamefont
  {Henny}}, \bibinfo {author} {\bibfnamefont {S.}~\bibnamefont {Oberholzer}},
  \bibinfo {author} {\bibfnamefont {C.}~\bibnamefont {Strunk}}, \bibinfo
  {author} {\bibfnamefont {T.}~\bibnamefont {Heinzel}}, \bibinfo {author}
  {\bibfnamefont {K.}~\bibnamefont {Ensslin}}, \bibinfo {author} {\bibfnamefont
  {M.}~\bibnamefont {Holland}},\ and\ \bibinfo {author} {\bibfnamefont
  {C.}~\bibnamefont {Sch{\"o}nenberger}},\ }\bibfield  {title} {\bibinfo
  {title} {The fermionic {H}anbury {B}rown and {T}wiss experiment},\
  }\href@noop {} {\bibfield  {journal} {\bibinfo  {journal} {Science}\ }\textbf
  {\bibinfo {volume} {284}},\ \bibinfo {pages} {296} (\bibinfo {year}
  {1999})}\BibitemShut {NoStop}%
\bibitem [{\citenamefont {Oliver}\ \emph {et~al.}(1999)\citenamefont {Oliver},
  \citenamefont {Kim}, \citenamefont {Liu},\ and\ \citenamefont
  {Yamamoto}}]{Oliver:1999}%
  \BibitemOpen
  \bibfield  {author} {\bibinfo {author} {\bibfnamefont {W.~D.}\ \bibnamefont
  {Oliver}}, \bibinfo {author} {\bibfnamefont {J.}~\bibnamefont {Kim}},
  \bibinfo {author} {\bibfnamefont {R.~C.}\ \bibnamefont {Liu}},\ and\ \bibinfo
  {author} {\bibfnamefont {Y.}~\bibnamefont {Yamamoto}},\ }\bibfield  {title}
  {\bibinfo {title} {{H}anbury {B}rown and {T}wiss-type experiment with
  electrons},\ }\href@noop {} {\bibfield  {journal} {\bibinfo  {journal}
  {Science}\ }\textbf {\bibinfo {volume} {284}},\ \bibinfo {pages} {299}
  (\bibinfo {year} {1999})}\BibitemShut {NoStop}%
\bibitem [{\citenamefont {Bocquillon}\ \emph {et~al.}(2013)\citenamefont
  {Bocquillon}, \citenamefont {Freulon}, \citenamefont {Berroir}, \citenamefont
  {Degiovanni}, \citenamefont {Pla{\c c}ais}, \citenamefont {Cavanna},
  \citenamefont {Jin},\ and\ \citenamefont {F{\`e}ve}}]{Bocquillon:2013}%
  \BibitemOpen
  \bibfield  {author} {\bibinfo {author} {\bibfnamefont {E.}~\bibnamefont
  {Bocquillon}}, \bibinfo {author} {\bibfnamefont {V.}~\bibnamefont {Freulon}},
  \bibinfo {author} {\bibfnamefont {J.-M.}\ \bibnamefont {Berroir}}, \bibinfo
  {author} {\bibfnamefont {P.}~\bibnamefont {Degiovanni}}, \bibinfo {author}
  {\bibfnamefont {B.}~\bibnamefont {Pla{\c c}ais}}, \bibinfo {author}
  {\bibfnamefont {A.}~\bibnamefont {Cavanna}}, \bibinfo {author} {\bibfnamefont
  {Y.}~\bibnamefont {Jin}},\ and\ \bibinfo {author} {\bibfnamefont
  {G.}~\bibnamefont {F{\`e}ve}},\ }\bibfield  {title} {\bibinfo {title}
  {Coherence and indistinguishability of single electrons emitted by
  independent sources},\ }\href@noop {} {\bibfield  {journal} {\bibinfo
  {journal} {Science}\ }\textbf {\bibinfo {volume} {339}},\ \bibinfo {pages}
  {1054} (\bibinfo {year} {2013})}\BibitemShut {NoStop}%
\bibitem [{\citenamefont {Kiesel}\ \emph {et~al.}(2002)\citenamefont {Kiesel},
  \citenamefont {Renz},\ and\ \citenamefont {Hasselbach}}]{Kiesel:2002}%
  \BibitemOpen
  \bibfield  {author} {\bibinfo {author} {\bibfnamefont {H.}~\bibnamefont
  {Kiesel}}, \bibinfo {author} {\bibfnamefont {A.}~\bibnamefont {Renz}},\ and\
  \bibinfo {author} {\bibfnamefont {F.}~\bibnamefont {Hasselbach}},\ }\bibfield
   {title} {\bibinfo {title} {Observation of {H}anbury {B}rown--{T}wiss
  anticorrelations for free electrons},\ }\href@noop {} {\bibfield  {journal}
  {\bibinfo  {journal} {Nature}\ }\textbf {\bibinfo {volume} {418}},\ \bibinfo
  {pages} {392} (\bibinfo {year} {2002})}\BibitemShut {NoStop}%
\bibitem [{\citenamefont {Lougovski}\ and\ \citenamefont
  {Batelaan}(2011)}]{Lougovski:2011}%
  \BibitemOpen
  \bibfield  {author} {\bibinfo {author} {\bibfnamefont {P.}~\bibnamefont
  {Lougovski}}\ and\ \bibinfo {author} {\bibfnamefont {H.}~\bibnamefont
  {Batelaan}},\ }\bibfield  {title} {\bibinfo {title} {Quantum description and
  properties of electrons emitted from pulsed nanotip electron sources},\
  }\href@noop {} {\bibfield  {journal} {\bibinfo  {journal} {Phys. Rev. A}\
  }\textbf {\bibinfo {volume} {84}},\ \bibinfo {pages} {023417} (\bibinfo
  {year} {2011})}\BibitemShut {NoStop}%
\bibitem [{\citenamefont {Baym}\ and\ \citenamefont {Shen}(2014)}]{Baym:2014}%
  \BibitemOpen
  \bibfield  {author} {\bibinfo {author} {\bibfnamefont {G.}~\bibnamefont
  {Baym}}\ and\ \bibinfo {author} {\bibfnamefont {K.}~\bibnamefont {Shen}},\
  }\bibinfo {title} {{H}anbury {B}rown--{T}wiss interferometry with electrons:
  Coulomb vs. quantum statistics}\ (\bibinfo  {publisher} {World Scientific},\
  \bibinfo {year} {2014})\ pp.\ \bibinfo {pages} {201--210}\BibitemShut
  {NoStop}%
\bibitem [{\citenamefont {Kuwahara}\ \emph
  {et~al.}(2021{\natexlab{a}})\citenamefont {Kuwahara}, \citenamefont
  {Yoshida}, \citenamefont {Nagata}, \citenamefont {Nakakura}, \citenamefont
  {Furui}, \citenamefont {Ishida}, \citenamefont {Saitoh}, \citenamefont
  {Ujihara},\ and\ \citenamefont {Tanaka}}]{Kuwahara:2021}%
  \BibitemOpen
  \bibfield  {author} {\bibinfo {author} {\bibfnamefont {M.}~\bibnamefont
  {Kuwahara}}, \bibinfo {author} {\bibfnamefont {Y.}~\bibnamefont {Yoshida}},
  \bibinfo {author} {\bibfnamefont {W.}~\bibnamefont {Nagata}}, \bibinfo
  {author} {\bibfnamefont {K.}~\bibnamefont {Nakakura}}, \bibinfo {author}
  {\bibfnamefont {M.}~\bibnamefont {Furui}}, \bibinfo {author} {\bibfnamefont
  {T.}~\bibnamefont {Ishida}}, \bibinfo {author} {\bibfnamefont
  {K.}~\bibnamefont {Saitoh}}, \bibinfo {author} {\bibfnamefont
  {T.}~\bibnamefont {Ujihara}},\ and\ \bibinfo {author} {\bibfnamefont
  {N.}~\bibnamefont {Tanaka}},\ }\bibfield  {title} {\bibinfo {title}
  {Intensity interference in a coherent spin-polarized electron beam},\
  }\href@noop {} {\bibfield  {journal} {\bibinfo  {journal} {Phys. Rev. Lett.}\
  }\textbf {\bibinfo {volume} {126}},\ \bibinfo {pages} {125501} (\bibinfo
  {year} {2021}{\natexlab{a}})}\BibitemShut {NoStop}%
\bibitem [{\citenamefont {Batelaan}\ \emph {et~al.}(2021)\citenamefont
  {Batelaan}, \citenamefont {Keramati},\ and\ \citenamefont
  {Gay}}]{Batelaan:2021}%
  \BibitemOpen
  \bibfield  {author} {\bibinfo {author} {\bibfnamefont {H.}~\bibnamefont
  {Batelaan}}, \bibinfo {author} {\bibfnamefont {S.}~\bibnamefont {Keramati}},\
  and\ \bibinfo {author} {\bibfnamefont {T.~J.}\ \bibnamefont {Gay}},\
  }\bibfield  {title} {\bibinfo {title} {Comment on ``{I}ntensity interference
  in a coherent spin-polarized electron beam''},\ }\href@noop {} {\bibfield
  {journal} {\bibinfo  {journal} {Phys. Rev. Lett.}\ }\textbf {\bibinfo
  {volume} {127}},\ \bibinfo {pages} {229601} (\bibinfo {year}
  {2021})}\BibitemShut {NoStop}%
\bibitem [{\citenamefont {Kuwahara}\ \emph
  {et~al.}(2021{\natexlab{b}})\citenamefont {Kuwahara}, \citenamefont
  {Yoshida}, \citenamefont {Nagata}, \citenamefont {Nakakura}, \citenamefont
  {Furui}, \citenamefont {Ishida}, \citenamefont {Saitoh}, \citenamefont
  {Ujihara},\ and\ \citenamefont {Tanaka}}]{Kuwahara:2021a}%
  \BibitemOpen
  \bibfield  {author} {\bibinfo {author} {\bibfnamefont {M.}~\bibnamefont
  {Kuwahara}}, \bibinfo {author} {\bibfnamefont {Y.}~\bibnamefont {Yoshida}},
  \bibinfo {author} {\bibfnamefont {W.}~\bibnamefont {Nagata}}, \bibinfo
  {author} {\bibfnamefont {K.}~\bibnamefont {Nakakura}}, \bibinfo {author}
  {\bibfnamefont {M.}~\bibnamefont {Furui}}, \bibinfo {author} {\bibfnamefont
  {T.}~\bibnamefont {Ishida}}, \bibinfo {author} {\bibfnamefont
  {K.}~\bibnamefont {Saitoh}}, \bibinfo {author} {\bibfnamefont
  {T.}~\bibnamefont {Ujihara}},\ and\ \bibinfo {author} {\bibfnamefont
  {N.}~\bibnamefont {Tanaka}},\ }\bibfield  {title} {\bibinfo {title} {Kuwahara
  et al. reply:},\ }\href@noop {} {\bibfield  {journal} {\bibinfo  {journal}
  {Phys. Rev. Lett.}\ }\textbf {\bibinfo {volume} {127}},\ \bibinfo {pages}
  {229602} (\bibinfo {year} {2021}{\natexlab{b}})}\BibitemShut {NoStop}%
\bibitem [{\citenamefont {Hommelhoff}\ \emph {et~al.}(2006)\citenamefont
  {Hommelhoff}, \citenamefont {Sortais}, \citenamefont {Aghajani-Talesh},\ and\
  \citenamefont {Kasevich}}]{Hommelhoff:2006a}%
  \BibitemOpen
  \bibfield  {author} {\bibinfo {author} {\bibfnamefont {P.}~\bibnamefont
  {Hommelhoff}}, \bibinfo {author} {\bibfnamefont {Y.}~\bibnamefont {Sortais}},
  \bibinfo {author} {\bibfnamefont {A.}~\bibnamefont {Aghajani-Talesh}},\ and\
  \bibinfo {author} {\bibfnamefont {M.~A.}\ \bibnamefont {Kasevich}},\
  }\bibfield  {title} {\bibinfo {title} {Field emission tip as a nanometer
  source of free electron femtosecond pulses},\ }\href@noop {} {\bibfield
  {journal} {\bibinfo  {journal} {Phys. Rev. Lett.}\ }\textbf {\bibinfo
  {volume} {96}},\ \bibinfo {pages} {077401} (\bibinfo {year}
  {2006})}\BibitemShut {NoStop}%
\bibitem [{\citenamefont {Ropers}\ \emph {et~al.}(2007)\citenamefont {Ropers},
  \citenamefont {Solli}, \citenamefont {Schulz}, \citenamefont {Lienau},\ and\
  \citenamefont {Elsaesser}}]{Ropers:2007}%
  \BibitemOpen
  \bibfield  {author} {\bibinfo {author} {\bibfnamefont {C.}~\bibnamefont
  {Ropers}}, \bibinfo {author} {\bibfnamefont {D.~R.}\ \bibnamefont {Solli}},
  \bibinfo {author} {\bibfnamefont {C.~P.}\ \bibnamefont {Schulz}}, \bibinfo
  {author} {\bibfnamefont {C.}~\bibnamefont {Lienau}},\ and\ \bibinfo {author}
  {\bibfnamefont {T.}~\bibnamefont {Elsaesser}},\ }\bibfield  {title} {\bibinfo
  {title} {Localized multiphoton emission of femtosecond electron pulses from
  metal nanotips},\ }\href@noop {} {\bibfield  {journal} {\bibinfo  {journal}
  {Phys. Rev. Lett.}\ }\textbf {\bibinfo {volume} {98}},\ \bibinfo {pages}
  {043907} (\bibinfo {year} {2007})}\BibitemShut {NoStop}%
\bibitem [{\citenamefont {Barwick}\ \emph {et~al.}(2007)\citenamefont
  {Barwick}, \citenamefont {Corder}, \citenamefont {Strohaber}, \citenamefont
  {Chandler-Smith}, \citenamefont {Uiterwaal},\ and\ \citenamefont
  {Batelaan}}]{Barwick:2007}%
  \BibitemOpen
  \bibfield  {author} {\bibinfo {author} {\bibfnamefont {B.}~\bibnamefont
  {Barwick}}, \bibinfo {author} {\bibfnamefont {C.}~\bibnamefont {Corder}},
  \bibinfo {author} {\bibfnamefont {J.}~\bibnamefont {Strohaber}}, \bibinfo
  {author} {\bibfnamefont {N.}~\bibnamefont {Chandler-Smith}}, \bibinfo
  {author} {\bibfnamefont {C.}~\bibnamefont {Uiterwaal}},\ and\ \bibinfo
  {author} {\bibfnamefont {H.}~\bibnamefont {Batelaan}},\ }\bibfield  {title}
  {\bibinfo {title} {Laser-induced ultrafast electron emission from a field
  emission tip},\ }\href@noop {} {\bibfield  {journal} {\bibinfo  {journal}
  {New Journal of Physics}\ }\textbf {\bibinfo {volume} {9}},\ \bibinfo {pages}
  {142} (\bibinfo {year} {2007})}\BibitemShut {NoStop}%
\bibitem [{\citenamefont {Ehberger}\ \emph {et~al.}(2015)\citenamefont
  {Ehberger}, \citenamefont {Hammer}, \citenamefont {Eisele}, \citenamefont
  {Kr\"uger}, \citenamefont {Noe}, \citenamefont {H\"ogele},\ and\
  \citenamefont {Hommelhoff}}]{Ehberger:2015}%
  \BibitemOpen
  \bibfield  {author} {\bibinfo {author} {\bibfnamefont {D.}~\bibnamefont
  {Ehberger}}, \bibinfo {author} {\bibfnamefont {J.}~\bibnamefont {Hammer}},
  \bibinfo {author} {\bibfnamefont {M.}~\bibnamefont {Eisele}}, \bibinfo
  {author} {\bibfnamefont {M.}~\bibnamefont {Kr\"uger}}, \bibinfo {author}
  {\bibfnamefont {J.}~\bibnamefont {Noe}}, \bibinfo {author} {\bibfnamefont
  {A.}~\bibnamefont {H\"ogele}},\ and\ \bibinfo {author} {\bibfnamefont
  {P.}~\bibnamefont {Hommelhoff}},\ }\bibfield  {title} {\bibinfo {title}
  {Highly coherent electron beam from a laser-triggered tungsten needle tip},\
  }\href@noop {} {\bibfield  {journal} {\bibinfo  {journal} {Phys. Rev. Lett.}\
  }\textbf {\bibinfo {volume} {114}},\ \bibinfo {pages} {227601} (\bibinfo
  {year} {2015})}\BibitemShut {NoStop}%
\bibitem [{\citenamefont {Meier}\ \emph {et~al.}(2018)\citenamefont {Meier},
  \citenamefont {Higuchi}, \citenamefont {Nutz}, \citenamefont {H{\"o}gele},\
  and\ \citenamefont {Hommelhoff}}]{Meier:2018}%
  \BibitemOpen
  \bibfield  {author} {\bibinfo {author} {\bibfnamefont {S.}~\bibnamefont
  {Meier}}, \bibinfo {author} {\bibfnamefont {T.}~\bibnamefont {Higuchi}},
  \bibinfo {author} {\bibfnamefont {M.}~\bibnamefont {Nutz}}, \bibinfo {author}
  {\bibfnamefont {A.}~\bibnamefont {H{\"o}gele}},\ and\ \bibinfo {author}
  {\bibfnamefont {P.}~\bibnamefont {Hommelhoff}},\ }\bibfield  {title}
  {\bibinfo {title} {{High spatial coherence in multiphoton-photoemitted
  electron beams}},\ }\href@noop {} {\bibfield  {journal} {\bibinfo  {journal}
  {Applied Physics Letters}\ }\textbf {\bibinfo {volume} {113}},\ \bibinfo
  {pages} {143101} (\bibinfo {year} {2018})}\BibitemShut {NoStop}%
\bibitem [{\citenamefont {Keramati}\ \emph {et~al.}(2021)\citenamefont
  {Keramati}, \citenamefont {Brunner}, \citenamefont {Gay},\ and\ \citenamefont
  {Batelaan}}]{Keramati:2021}%
  \BibitemOpen
  \bibfield  {author} {\bibinfo {author} {\bibfnamefont {S.}~\bibnamefont
  {Keramati}}, \bibinfo {author} {\bibfnamefont {W.}~\bibnamefont {Brunner}},
  \bibinfo {author} {\bibfnamefont {T.~J.}\ \bibnamefont {Gay}},\ and\ \bibinfo
  {author} {\bibfnamefont {H.}~\bibnamefont {Batelaan}},\ }\bibfield  {title}
  {\bibinfo {title} {Non-poissonian ultrashort nanoscale electron pulses},\
  }\href@noop {} {\bibfield  {journal} {\bibinfo  {journal} {Phys. Rev. Lett.}\
  }\textbf {\bibinfo {volume} {127}},\ \bibinfo {pages} {180602} (\bibinfo
  {year} {2021})}\BibitemShut {NoStop}%
\bibitem [{\citenamefont {Haindl}\ \emph {et~al.}(2023)\citenamefont {Haindl},
  \citenamefont {Feist}, \citenamefont {Domr{\"o}se}, \citenamefont
  {M{\"o}ller}, \citenamefont {Gaida}, \citenamefont {Yalunin},\ and\
  \citenamefont {Ropers}}]{Haindl:2023}%
  \BibitemOpen
  \bibfield  {author} {\bibinfo {author} {\bibfnamefont {R.}~\bibnamefont
  {Haindl}}, \bibinfo {author} {\bibfnamefont {A.}~\bibnamefont {Feist}},
  \bibinfo {author} {\bibfnamefont {T.}~\bibnamefont {Domr{\"o}se}}, \bibinfo
  {author} {\bibfnamefont {M.}~\bibnamefont {M{\"o}ller}}, \bibinfo {author}
  {\bibfnamefont {J.~H.}\ \bibnamefont {Gaida}}, \bibinfo {author}
  {\bibfnamefont {S.~V.}\ \bibnamefont {Yalunin}},\ and\ \bibinfo {author}
  {\bibfnamefont {C.}~\bibnamefont {Ropers}},\ }\bibfield  {title} {\bibinfo
  {title} {Coulomb-correlated electron number states in a transmission electron
  microscope beam},\ }\href@noop {} {\bibfield  {journal} {\bibinfo  {journal}
  {Nature Physics}\ }\textbf {\bibinfo {volume} {19}},\ \bibinfo {pages} {1410}
  (\bibinfo {year} {2023})}\BibitemShut {NoStop}%
\bibitem [{\citenamefont {Meier}\ \emph {et~al.}(2023)\citenamefont {Meier},
  \citenamefont {Heimerl},\ and\ \citenamefont {Hommelhoff}}]{Meier:2023}%
  \BibitemOpen
  \bibfield  {author} {\bibinfo {author} {\bibfnamefont {S.}~\bibnamefont
  {Meier}}, \bibinfo {author} {\bibfnamefont {J.}~\bibnamefont {Heimerl}},\
  and\ \bibinfo {author} {\bibfnamefont {P.}~\bibnamefont {Hommelhoff}},\
  }\bibfield  {title} {\bibinfo {title} {Few-electron correlations after
  ultrafast photoemission from nanometric needle tips},\ }\href@noop {}
  {\bibfield  {journal} {\bibinfo  {journal} {Nature Physics}\ }\textbf
  {\bibinfo {volume} {19}},\ \bibinfo {pages} {1402} (\bibinfo {year}
  {2023})}\BibitemShut {NoStop}%
\bibitem [{\citenamefont {Zou}\ \emph {et~al.}(1991)\citenamefont {Zou},
  \citenamefont {Wang},\ and\ \citenamefont {Mandel}}]{Zou:1991}%
  \BibitemOpen
  \bibfield  {author} {\bibinfo {author} {\bibfnamefont {X.~Y.}\ \bibnamefont
  {Zou}}, \bibinfo {author} {\bibfnamefont {L.~J.}\ \bibnamefont {Wang}},\ and\
  \bibinfo {author} {\bibfnamefont {L.}~\bibnamefont {Mandel}},\ }\bibfield
  {title} {\bibinfo {title} {Induced coherence and indistinguishability in
  optical interference},\ }\href@noop {} {\bibfield  {journal} {\bibinfo
  {journal} {Phys. Rev. Lett.}\ }\textbf {\bibinfo {volume} {67}},\ \bibinfo
  {pages} {318} (\bibinfo {year} {1991})}\BibitemShut {NoStop}%
\bibitem [{\citenamefont {Lemos}\ \emph {et~al.}(2014)\citenamefont {Lemos},
  \citenamefont {Borish}, \citenamefont {Cole}, \citenamefont {Ramelow},
  \citenamefont {Lapkiewicz},\ and\ \citenamefont {Zeilinger}}]{Lemos:2014}%
  \BibitemOpen
  \bibfield  {author} {\bibinfo {author} {\bibfnamefont {G.~B.}\ \bibnamefont
  {Lemos}}, \bibinfo {author} {\bibfnamefont {V.}~\bibnamefont {Borish}},
  \bibinfo {author} {\bibfnamefont {G.~D.}\ \bibnamefont {Cole}}, \bibinfo
  {author} {\bibfnamefont {S.}~\bibnamefont {Ramelow}}, \bibinfo {author}
  {\bibfnamefont {R.}~\bibnamefont {Lapkiewicz}},\ and\ \bibinfo {author}
  {\bibfnamefont {A.}~\bibnamefont {Zeilinger}},\ }\bibfield  {title} {\bibinfo
  {title} {Quantum imaging with undetected photons},\ }\href@noop {} {\bibfield
   {journal} {\bibinfo  {journal} {Nature}\ }\textbf {\bibinfo {volume}
  {512}},\ \bibinfo {pages} {409} (\bibinfo {year} {2014})}\BibitemShut
  {NoStop}%
\bibitem [{\citenamefont {Oppel}\ \emph {et~al.}(2012)\citenamefont {Oppel},
  \citenamefont {B\"uttner}, \citenamefont {Kok},\ and\ \citenamefont {von
  Zanthier}}]{Oppel:2012}%
  \BibitemOpen
  \bibfield  {author} {\bibinfo {author} {\bibfnamefont {S.}~\bibnamefont
  {Oppel}}, \bibinfo {author} {\bibfnamefont {T.}~\bibnamefont {B\"uttner}},
  \bibinfo {author} {\bibfnamefont {P.}~\bibnamefont {Kok}},\ and\ \bibinfo
  {author} {\bibfnamefont {J.}~\bibnamefont {von Zanthier}},\ }\bibfield
  {title} {\bibinfo {title} {Superresolving multiphoton interferences with
  independent light sources},\ }\href@noop {} {\bibfield  {journal} {\bibinfo
  {journal} {Phys. Rev. Lett.}\ }\textbf {\bibinfo {volume} {109}},\ \bibinfo
  {pages} {233603} (\bibinfo {year} {2012})}\BibitemShut {NoStop}%
\bibitem [{\citenamefont {Classen}\ \emph {et~al.}(2016)\citenamefont
  {Classen}, \citenamefont {Waldmann}, \citenamefont {Giebel}, \citenamefont
  {Schneider}, \citenamefont {Bhatti}, \citenamefont {Mehringer},\ and\
  \citenamefont {von Zanthier}}]{Classen:2016}%
  \BibitemOpen
  \bibfield  {author} {\bibinfo {author} {\bibfnamefont {A.}~\bibnamefont
  {Classen}}, \bibinfo {author} {\bibfnamefont {F.}~\bibnamefont {Waldmann}},
  \bibinfo {author} {\bibfnamefont {S.}~\bibnamefont {Giebel}}, \bibinfo
  {author} {\bibfnamefont {R.}~\bibnamefont {Schneider}}, \bibinfo {author}
  {\bibfnamefont {D.}~\bibnamefont {Bhatti}}, \bibinfo {author} {\bibfnamefont
  {T.}~\bibnamefont {Mehringer}},\ and\ \bibinfo {author} {\bibfnamefont
  {J.}~\bibnamefont {von Zanthier}},\ }\bibfield  {title} {\bibinfo {title}
  {Superresolving imaging of arbitrary one-dimensional arrays of thermal light
  sources using multiphoton interference},\ }\href@noop {} {\bibfield
  {journal} {\bibinfo  {journal} {Phys. Rev. Lett.}\ }\textbf {\bibinfo
  {volume} {117}},\ \bibinfo {pages} {253601} (\bibinfo {year}
  {2016})}\BibitemShut {NoStop}%
\bibitem [{\citenamefont {Classen}\ \emph {et~al.}(2017)\citenamefont
  {Classen}, \citenamefont {Ayyer}, \citenamefont {Chapman}, \citenamefont
  {R\"ohlsberger},\ and\ \citenamefont {von Zanthier}}]{Classen:2017}%
  \BibitemOpen
  \bibfield  {author} {\bibinfo {author} {\bibfnamefont {A.}~\bibnamefont
  {Classen}}, \bibinfo {author} {\bibfnamefont {K.}~\bibnamefont {Ayyer}},
  \bibinfo {author} {\bibfnamefont {H.~N.}\ \bibnamefont {Chapman}}, \bibinfo
  {author} {\bibfnamefont {R.}~\bibnamefont {R\"ohlsberger}},\ and\ \bibinfo
  {author} {\bibfnamefont {J.}~\bibnamefont {von Zanthier}},\ }\bibfield
  {title} {\bibinfo {title} {Incoherent diffractive imaging via intensity
  correlations of hard x rays},\ }\href@noop {} {\bibfield  {journal} {\bibinfo
   {journal} {Phys. Rev. Lett.}\ }\textbf {\bibinfo {volume} {119}},\ \bibinfo
  {pages} {053401} (\bibinfo {year} {2017})}\BibitemShut {NoStop}%
\bibitem [{\citenamefont {Schneider}\ \emph {et~al.}(2018)\citenamefont
  {Schneider}, \citenamefont {Mehringer}, \citenamefont {Mercurio},
  \citenamefont {Wenthaus}, \citenamefont {Classen}, \citenamefont {Brenner},
  \citenamefont {Gorobtsov}, \citenamefont {Benz}, \citenamefont {Bhatti},
  \citenamefont {Bocklage}, \citenamefont {Fischer}, \citenamefont {Lazarev},
  \citenamefont {Obukhov}, \citenamefont {Schlage}, \citenamefont {Skopintsev},
  \citenamefont {Wagner}, \citenamefont {Waldmann}, \citenamefont {Willing},
  \citenamefont {Zaluzhnyy}, \citenamefont {Wurth}, \citenamefont
  {Vartanyants}, \citenamefont {R{\"o}hlsberger},\ and\ \citenamefont {von
  Zanthier}}]{Schneider:2018}%
  \BibitemOpen
  \bibfield  {author} {\bibinfo {author} {\bibfnamefont {R.}~\bibnamefont
  {Schneider}}, \bibinfo {author} {\bibfnamefont {T.}~\bibnamefont
  {Mehringer}}, \bibinfo {author} {\bibfnamefont {G.}~\bibnamefont {Mercurio}},
  \bibinfo {author} {\bibfnamefont {L.}~\bibnamefont {Wenthaus}}, \bibinfo
  {author} {\bibfnamefont {A.}~\bibnamefont {Classen}}, \bibinfo {author}
  {\bibfnamefont {G.}~\bibnamefont {Brenner}}, \bibinfo {author} {\bibfnamefont
  {O.}~\bibnamefont {Gorobtsov}}, \bibinfo {author} {\bibfnamefont
  {A.}~\bibnamefont {Benz}}, \bibinfo {author} {\bibfnamefont {D.}~\bibnamefont
  {Bhatti}}, \bibinfo {author} {\bibfnamefont {L.}~\bibnamefont {Bocklage}},
  \bibinfo {author} {\bibfnamefont {B.}~\bibnamefont {Fischer}}, \bibinfo
  {author} {\bibfnamefont {S.}~\bibnamefont {Lazarev}}, \bibinfo {author}
  {\bibfnamefont {Y.}~\bibnamefont {Obukhov}}, \bibinfo {author} {\bibfnamefont
  {K.}~\bibnamefont {Schlage}}, \bibinfo {author} {\bibfnamefont
  {P.}~\bibnamefont {Skopintsev}}, \bibinfo {author} {\bibfnamefont
  {J.}~\bibnamefont {Wagner}}, \bibinfo {author} {\bibfnamefont
  {F.}~\bibnamefont {Waldmann}}, \bibinfo {author} {\bibfnamefont
  {S.}~\bibnamefont {Willing}}, \bibinfo {author} {\bibfnamefont
  {I.}~\bibnamefont {Zaluzhnyy}}, \bibinfo {author} {\bibfnamefont
  {W.}~\bibnamefont {Wurth}}, \bibinfo {author} {\bibfnamefont {I.~A.}\
  \bibnamefont {Vartanyants}}, \bibinfo {author} {\bibfnamefont
  {R.}~\bibnamefont {R{\"o}hlsberger}},\ and\ \bibinfo {author} {\bibfnamefont
  {J.}~\bibnamefont {von Zanthier}},\ }\bibfield  {title} {\bibinfo {title}
  {Quantum imaging with incoherently scattered light from a free-electron
  laser},\ }\href@noop {} {\bibfield  {journal} {\bibinfo  {journal} {Nature
  Physics}\ }\textbf {\bibinfo {volume} {14}},\ \bibinfo {pages} {126}
  (\bibinfo {year} {2018})}\BibitemShut {NoStop}%
\bibitem [{\citenamefont {Ho}\ \emph {et~al.}(2021)\citenamefont {Ho},
  \citenamefont {Knight},\ and\ \citenamefont {Young}}]{Ho:2021}%
  \BibitemOpen
  \bibfield  {author} {\bibinfo {author} {\bibfnamefont {P.~J.}\ \bibnamefont
  {Ho}}, \bibinfo {author} {\bibfnamefont {C.}~\bibnamefont {Knight}},\ and\
  \bibinfo {author} {\bibfnamefont {L.}~\bibnamefont {Young}},\ }\bibfield
  {title} {\bibinfo {title} {{Fluorescence intensity correlation imaging with
  high spatial resolution and elemental contrast using intense x-ray pulses}},\
  }\href@noop {} {\bibfield  {journal} {\bibinfo  {journal} {Structural
  Dynamics}\ }\textbf {\bibinfo {volume} {8}},\ \bibinfo {pages} {044101}
  (\bibinfo {year} {2021})}\BibitemShut {NoStop}%
\bibitem [{\citenamefont {Richter}\ \emph {et~al.}(2021)\citenamefont
  {Richter}, \citenamefont {Wolf}, \citenamefont {von Zanthier},\ and\
  \citenamefont {Schmidt-Kaler}}]{Richter:2021}%
  \BibitemOpen
  \bibfield  {author} {\bibinfo {author} {\bibfnamefont {S.}~\bibnamefont
  {Richter}}, \bibinfo {author} {\bibfnamefont {S.}~\bibnamefont {Wolf}},
  \bibinfo {author} {\bibfnamefont {J.}~\bibnamefont {von Zanthier}},\ and\
  \bibinfo {author} {\bibfnamefont {F.}~\bibnamefont {Schmidt-Kaler}},\
  }\bibfield  {title} {\bibinfo {title} {Imaging trapped ion structures via
  fluorescence cross-correlation detection},\ }\href@noop {} {\bibfield
  {journal} {\bibinfo  {journal} {Phys. Rev. Lett.}\ }\textbf {\bibinfo
  {volume} {126}},\ \bibinfo {pages} {173602} (\bibinfo {year}
  {2021})}\BibitemShut {NoStop}%
\bibitem [{\citenamefont {Trost}\ \emph {et~al.}(2023)\citenamefont {Trost},
  \citenamefont {Ayyer}, \citenamefont {Prasciolu}, \citenamefont
  {Fleckenstein}, \citenamefont {Barthelmess}, \citenamefont {Yefanov},
  \citenamefont {Dresselhaus}, \citenamefont {Li}, \citenamefont {Bajt},
  \citenamefont {Carnis}, \citenamefont {Wollweber}, \citenamefont {Mall},
  \citenamefont {Shen}, \citenamefont {Zhuang}, \citenamefont {Richter},
  \citenamefont {Karl}, \citenamefont {Cardoch}, \citenamefont {Patra},
  \citenamefont {M\"oller}, \citenamefont {Zozulya}, \citenamefont {Shayduk},
  \citenamefont {Lu}, \citenamefont {Brau\ss{}e}, \citenamefont {Friedrich},
  \citenamefont {Boesenberg}, \citenamefont {Petrov}, \citenamefont {Tomin},
  \citenamefont {Guetg}, \citenamefont {Madsen}, \citenamefont {Timneanu},
  \citenamefont {Caleman}, \citenamefont {R\"ohlsberger}, \citenamefont {von
  Zanthier},\ and\ \citenamefont {Chapman}}]{Trost:2023}%
  \BibitemOpen
  \bibfield  {author} {\bibinfo {author} {\bibfnamefont {F.}~\bibnamefont
  {Trost}}, \bibinfo {author} {\bibfnamefont {K.}~\bibnamefont {Ayyer}},
  \bibinfo {author} {\bibfnamefont {M.}~\bibnamefont {Prasciolu}}, \bibinfo
  {author} {\bibfnamefont {H.}~\bibnamefont {Fleckenstein}}, \bibinfo {author}
  {\bibfnamefont {M.}~\bibnamefont {Barthelmess}}, \bibinfo {author}
  {\bibfnamefont {O.}~\bibnamefont {Yefanov}}, \bibinfo {author} {\bibfnamefont
  {J.~L.}\ \bibnamefont {Dresselhaus}}, \bibinfo {author} {\bibfnamefont
  {C.}~\bibnamefont {Li}}, \bibinfo {author} {\bibfnamefont {S.~c.~v.}\
  \bibnamefont {Bajt}}, \bibinfo {author} {\bibfnamefont {J.}~\bibnamefont
  {Carnis}}, \bibinfo {author} {\bibfnamefont {T.}~\bibnamefont {Wollweber}},
  \bibinfo {author} {\bibfnamefont {A.}~\bibnamefont {Mall}}, \bibinfo {author}
  {\bibfnamefont {Z.}~\bibnamefont {Shen}}, \bibinfo {author} {\bibfnamefont
  {Y.}~\bibnamefont {Zhuang}}, \bibinfo {author} {\bibfnamefont
  {S.}~\bibnamefont {Richter}}, \bibinfo {author} {\bibfnamefont
  {S.}~\bibnamefont {Karl}}, \bibinfo {author} {\bibfnamefont {S.}~\bibnamefont
  {Cardoch}}, \bibinfo {author} {\bibfnamefont {K.~K.}\ \bibnamefont {Patra}},
  \bibinfo {author} {\bibfnamefont {J.}~\bibnamefont {M\"oller}}, \bibinfo
  {author} {\bibfnamefont {A.}~\bibnamefont {Zozulya}}, \bibinfo {author}
  {\bibfnamefont {R.}~\bibnamefont {Shayduk}}, \bibinfo {author} {\bibfnamefont
  {W.}~\bibnamefont {Lu}}, \bibinfo {author} {\bibfnamefont {F.}~\bibnamefont
  {Brau\ss{}e}}, \bibinfo {author} {\bibfnamefont {B.}~\bibnamefont
  {Friedrich}}, \bibinfo {author} {\bibfnamefont {U.}~\bibnamefont
  {Boesenberg}}, \bibinfo {author} {\bibfnamefont {I.}~\bibnamefont {Petrov}},
  \bibinfo {author} {\bibfnamefont {S.}~\bibnamefont {Tomin}}, \bibinfo
  {author} {\bibfnamefont {M.}~\bibnamefont {Guetg}}, \bibinfo {author}
  {\bibfnamefont {A.}~\bibnamefont {Madsen}}, \bibinfo {author} {\bibfnamefont
  {N.}~\bibnamefont {Timneanu}}, \bibinfo {author} {\bibfnamefont
  {C.}~\bibnamefont {Caleman}}, \bibinfo {author} {\bibfnamefont
  {R.}~\bibnamefont {R\"ohlsberger}}, \bibinfo {author} {\bibfnamefont
  {J.}~\bibnamefont {von Zanthier}},\ and\ \bibinfo {author} {\bibfnamefont
  {H.~N.}\ \bibnamefont {Chapman}},\ }\bibfield  {title} {\bibinfo {title}
  {Imaging via correlation of x-ray fluorescence photons},\ }\href@noop {}
  {\bibfield  {journal} {\bibinfo  {journal} {Phys. Rev. Lett.}\ }\textbf
  {\bibinfo {volume} {130}},\ \bibinfo {pages} {173201} (\bibinfo {year}
  {2023})}\BibitemShut {NoStop}%
\bibitem [{Note1()}]{Note1}%
  \BibitemOpen
  \bibinfo {note} {In \protect \textit {spatial} second-order correlations, an
  energy uncertainty of around $1\%$ will merely lead to a slight reduction of
  the contrast.}\BibitemShut {Stop}%
\bibitem [{\citenamefont {Skornia}\ \emph {et~al.}(2001)\citenamefont
  {Skornia}, \citenamefont {von Zanthier}, \citenamefont {Agarwal},
  \citenamefont {Werner},\ and\ \citenamefont {Walther}}]{Skornia:2001}%
  \BibitemOpen
  \bibfield  {author} {\bibinfo {author} {\bibfnamefont {C.}~\bibnamefont
  {Skornia}}, \bibinfo {author} {\bibfnamefont {J.}~\bibnamefont {von
  Zanthier}}, \bibinfo {author} {\bibfnamefont {G.~S.}\ \bibnamefont
  {Agarwal}}, \bibinfo {author} {\bibfnamefont {E.}~\bibnamefont {Werner}},\
  and\ \bibinfo {author} {\bibfnamefont {H.}~\bibnamefont {Walther}},\
  }\bibfield  {title} {\bibinfo {title} {Nonclassical interference effects in
  the radiation from coherently driven uncorrelated atoms},\ }\href@noop {}
  {\bibfield  {journal} {\bibinfo  {journal} {Phys. Rev. A}\ }\textbf {\bibinfo
  {volume} {64}},\ \bibinfo {pages} {063801} (\bibinfo {year}
  {2001})}\BibitemShut {NoStop}%
\bibitem [{\citenamefont {Thiel}\ \emph {et~al.}(2007)\citenamefont {Thiel},
  \citenamefont {Bastin}, \citenamefont {Martin}, \citenamefont {Solano},
  \citenamefont {von Zanthier},\ and\ \citenamefont {Agarwal}}]{Thiel:2007}%
  \BibitemOpen
  \bibfield  {author} {\bibinfo {author} {\bibfnamefont {C.}~\bibnamefont
  {Thiel}}, \bibinfo {author} {\bibfnamefont {T.}~\bibnamefont {Bastin}},
  \bibinfo {author} {\bibfnamefont {J.}~\bibnamefont {Martin}}, \bibinfo
  {author} {\bibfnamefont {E.}~\bibnamefont {Solano}}, \bibinfo {author}
  {\bibfnamefont {J.}~\bibnamefont {von Zanthier}},\ and\ \bibinfo {author}
  {\bibfnamefont {G.~S.}\ \bibnamefont {Agarwal}},\ }\bibfield  {title}
  {\bibinfo {title} {Quantum imaging with incoherent photons},\ }\href@noop {}
  {\bibfield  {journal} {\bibinfo  {journal} {Phys. Rev. Lett.}\ }\textbf
  {\bibinfo {volume} {99}},\ \bibinfo {pages} {133603} (\bibinfo {year}
  {2007})}\BibitemShut {NoStop}%
\bibitem [{\citenamefont {M{\"a}hrlein}\ \emph {et~al.}(2017)\citenamefont
  {M{\"a}hrlein}, \citenamefont {Oppel}, \citenamefont {Wiegner},\ and\
  \citenamefont {von Zanthier}}]{Mahrlein:2017}%
  \BibitemOpen
  \bibfield  {author} {\bibinfo {author} {\bibfnamefont {S.}~\bibnamefont
  {M{\"a}hrlein}}, \bibinfo {author} {\bibfnamefont {S.}~\bibnamefont {Oppel}},
  \bibinfo {author} {\bibfnamefont {R.}~\bibnamefont {Wiegner}},\ and\ \bibinfo
  {author} {\bibfnamefont {J.}~\bibnamefont {von Zanthier}},\ }\bibfield
  {title} {\bibinfo {title} {Hong--{Ou}--{M}andel interference without beam
  splitters},\ }\href@noop {} {\bibfield  {journal} {\bibinfo  {journal}
  {Journal of Modern Optics}\ }\textbf {\bibinfo {volume} {64}},\ \bibinfo
  {pages} {921} (\bibinfo {year} {2017})}\BibitemShut {NoStop}%
\bibitem [{Note2()}]{Note2}%
  \BibitemOpen
  \bibinfo {note} {Note that any random emission phases cancel in Eq.~\protect
  \eqref {ggeral} due to the joint appearance of creation and annihilation
  operator for each source.}\BibitemShut {Stop}%
\bibitem [{\citenamefont {Mandel}(1983)}]{Mandel:1983}%
  \BibitemOpen
  \bibfield  {author} {\bibinfo {author} {\bibfnamefont {L.}~\bibnamefont
  {Mandel}},\ }\bibfield  {title} {\bibinfo {title} {Photon interference and
  correlation effects produced by independent quantum sources},\ }\href@noop {}
  {\bibfield  {journal} {\bibinfo  {journal} {Phys. Rev. A}\ }\textbf {\bibinfo
  {volume} {28}},\ \bibinfo {pages} {929} (\bibinfo {year} {1983})}\BibitemShut
  {NoStop}%
\bibitem [{\citenamefont {Verlet}(1967)}]{Verlet:1967}%
  \BibitemOpen
  \bibfield  {author} {\bibinfo {author} {\bibfnamefont {L.}~\bibnamefont
  {Verlet}},\ }\bibfield  {title} {\bibinfo {title} {Computer "experiments" on
  classical fluids. {I}. {T}hermodynamical properties of lennard-jones
  molecules},\ }\href@noop {} {\bibfield  {journal} {\bibinfo  {journal} {Phys.
  Rev.}\ }\textbf {\bibinfo {volume} {159}},\ \bibinfo {pages} {98} (\bibinfo
  {year} {1967})}\BibitemShut {NoStop}%
\end{thebibliography}%

\end{document}